\documentclass[twocolumn,journal]{IEEEtran}

\IEEEoverridecommandlockouts

\usepackage{color}
\usepackage{graphicx}
\usepackage{epstopdf}
\usepackage{amsmath}
\usepackage{amssymb}
\usepackage{algorithm}
\usepackage{algorithmic}
\usepackage{amsmath}
\usepackage{multirow}
\usepackage{booktabs}
\usepackage{array}
\usepackage{amsthm}
\usepackage{stfloats}
\usepackage{caption}
\usepackage{subfigure}
\usepackage{bm}
\usepackage{setspace}
\usepackage{mdframed}
\usepackage{diagbox}

\newcommand{\be}{\begin{equation}}
\newcommand{\ee}{\end{equation}}
\newcommand{\bea}{\begin{eqnarray}}
\newcommand{\eea}{\end{eqnarray}}
\newcommand{\ba}{\begin{array}}
\newcommand{\ea}{\end{array}}
\newcommand{\nid}{\noindent}

\title{Joint Sensing and Communication Optimization in Target-Mounted STARS-Assisted Vehicular Networks: A MADRL Approach
\thanks{H. Zhang, M. Li, and W. Wang are with the School of Information and Communication Engineering, Dalian University of Technology, Dalian 116024, China (e-mail: dlutzhc@mail.dlut.edu.cn; mli@dlut.edu.cn; wangwei2023@dlut.edu.cn).}
\thanks{R. Liu is with the Center for Pervasive Communications and Computing, University of California, Irvine, CA 92697, USA (e-mail: rangl2@uci.edu).}
\thanks{Q. Liu is with the School of Computer Science and Technology, Dalian University of Technology, Dalian 116024, China (e-mail: qianliu@dlut.edu.cn).}}
\author{Haocheng Zhang,
        Rang Liu,~\IEEEmembership{Graduate Student Member,~IEEE,}
        Ming Li,~\IEEEmembership{Senior Member,~IEEE,}\\
        Wei Wang,~\IEEEmembership{Member,~IEEE,}
        and Qian Liu,~\IEEEmembership{Member,~IEEE}
}

\begin{document}
\maketitle
\pagestyle{empty}
\thispagestyle{empty}


\begin{abstract}
The utilization of integrated sensing and communication (ISAC) technology has the potential to enhance the communication performance of road side units (RSUs) through the active sensing of target vehicles.
Furthermore, installing a simultaneous transmitting and reflecting surface (STARS) on the target vehicle can provide an extra boost to the reflection of the echo signal, thereby improving the communication quality for in-vehicle users.
However, the design of this target-mounted STARS system exhibits significant challenges, such as limited information sharing and distributed STARS control.
In this paper, we propose an end-to-end multi-agent deep reinforcement learning (MADRL) framework to tackle the challenges of joint sensing and communication optimization in the considered target-mounted STARS assisted vehicle networks.
By deploying agents on both RSU and vehicle, the MADRL framework enables RSU and vehicle to perform beam prediction and STARS pre-configuration using their respective local information.
To ensure efficient and stable learning for continuous decision-making, we employ the multi-agent soft actor critic (MASAC) algorithm and the multi-agent proximal policy optimization (MAPPO) algorithm on the proposed MADRL framework.
Extensive experimental results confirm the effectiveness of our proposed MADRL framework in improving both sensing and communication performance through the utilization of target-mounted STARS.
Finally, we conduct a comparative analysis and comparison of the two proposed algorithms under various environmental conditions.
\end{abstract}


\section{Introduction}

Sensing capabilities will play a crucial role in the sixth-generation (6G) wireless networks \cite{fliu}. The demand for higher-resolution localization motivates the development of environment-aware technologies including vehicle-to-everything (V2X) and virtual reality (VR) \cite{cxwang}. Meanwhile, accurate sensing ability presents an opportunity to improve the quality of service (QoS) in communications. \cite{yhcui}. Moreover, the potential key technologies in 6G, such as in-band full-duplex (IBFD) and ultra-massive multiple-input-multiple-output (MIMO), provide new chances to further facilitate mutual assistance between sensing and communication (S\&C). Through the utilization of shared hardware and spectrum resources, integrated sensing and communication (ISAC) exploits coordination gains to achieve better resource management and improved efficiency of S\&C \cite{fwdong}.

Vehicular networks are one of the most important application scenarios of ISAC. With the leaping development of autonomous driving and intelligent transportation, vehicles of the next generation require stronger self-awareness and environmental awareness abilities. ISAC systems allow for large-scale multiview sensing data sharing among vehicles and infrastructure, improving the reliability and efficiency of transportation systems \cite{qxzhang}-\cite{pxliu}. Besides, ISAC systems play an important role in sensing-assisted beamforming design for vehicle-to-infrastructure (V2I) communications. Specifically, the road side unit (RSU) directly utilizes echo signals reflected by the vehicles to predict beams for data transmission, which avoids high signaling overhead and frequent feedback. In order to achieve more accurate beam prediction, the researchers implement a variety of techniques for processing reflected signals to more precisely detect and track vehicles, such as extended Kalman filter (EKF), factor graphs and deep learning (DL) techniques \cite{fliu2}-\cite{zhwang}.

Considering the dynamic electromagnetic (EM) conditions prevalent in vehicular networks, there is a growing trend towards adopting reconfigurable intelligent surfaces (RIS) to improve signal propagation \cite{mnar}, \cite{rliu}.
RIS is a meta-surface consisting of massive EM elements, each of which can intelligently adjust the parameters of incident signals \cite{qqwu}. Specifically in vehicular networks, RIS can efficiently combat high path loss in high-frequency bands to enhance V2X connectivity and achieve capacity gains through shaping wireless environment \cite{ybchen}, \cite{pzhang}. The authors in \cite{ybchen2} demonstrated the significant role of RIS in improving the QoS performance of V2I communications. The authors in \cite{yai} validated the enhancement of confidentiality in V2X communications by RIS. In \cite{ab}, the authors conducted an in-depth investigation into RIS-enabled unmanned aerial vehicle (UAV)-based vehicular communication networks. However, RIS can only provide services for half-space where both the source and destination nodes lie on the same side of the RIS \cite{qqwu2}. To tackle this issue, the simultaneous transmitting and reflecting surface (STARS, a.k.a., STAR-RIS) is designed to support reconfiguring the transmitted and reflected signals via transmission and reflection coefficients, significantly increasing the degrees of freedom (DoFs) in signal propagation manipulation \cite{ywliu}.
On the other hand, the additional DoFs offered by STARS may potentially contribute to balancing the requirements of S\&C that are inherently conflicting in ISAC systems.


Currently, research on the STARS (or RIS)-assisted ISAC vehicular networks is in the initial stages. The majority of prior research has been carried out in static scenarios where S\&C objectives are distinct \cite{xnliu}, \cite{zlwang}.
To maximize the role of STARS in 6G-V2X systems, a promising approach is to mount STARS on the surface of target vehicles to improve S\&C performance. On the one hand, benefiting from the reflection function of target-mounted STARS, the sensing capabilities of RSUs can be effectively enhanced by increasing the radar cross-section (RCS) of the target vehicle \cite{plwang}-\cite{xdshao}.
On the other hand, despite suffering from high loss when high-frequency signals penetrate the target vehicle, target-mounted STARS have the capability to improve the communication performance of in-vehicle users through the refraction function \cite{zxhuang}.
More importantly, by appropriately leveraging both the reflection and refraction capabilities of STARS, it has the potential to enhance the effectiveness of sensing-assisted communications, thereby ultimately achieving superior communication quality for the target-mounted STARS vehicular networks \cite{ktmeng}.

The above-mentioned works are of significant importance for the deployment of target-mounted STARS. However, they have certain limitations. Firstly, the control of target-mounted STARS should be realized in the vehicles rather than the RSU. Prior work \cite{plwang}-\cite{zxhuang} assumed that RSUs control and configure target-mounted STARS, which results in complex transmission protocol designs, lower reliability and security problems. Secondly, in the non-stationary and time-varying environment of 6G-V2X, both the RSU and the vehicle can only observe localized environmental information. Frequent information transmission (especially uplink communication) only designed for sharing information is unrealistic. Furthermore, for rapidly changing vehicular network environments, performing high-complexity optimization algorithms such as channel state information (CSI) estimation introduces significant signaling overhead. In the target-mounted STARS system, the RSU enhances the target sensing performance by relying on STARS reflective capabilities without the need for additional measurements. The research in \cite{zxhuang} conducts additional CSI estimation and the research in \cite{ktmeng} requires additional vehicle parameters measurement at the RSU when using EKF technology.

In this paper, we introduce a multi-agent deep reinforcement learning (MADRL) approach to deal with the challenges in the target-mounted STARS system, which is distinguished by distributed decision-making in a dynamic environment \cite{txli}. DRL is considered as a promising method to address physical layer optimization \cite{ncl}, \cite{zhxiong}, such as modulation, beamforming design and channel estimation \cite{cwhuang}-\cite{lzhang}. Compared with centralized processing, MADRL can compromise cooperative and competitive trade-offs of agents to achieve a flexible balance in V2I networks \cite{lliang}-\cite{xli}. Incorporating the advantages of MADRL into the target-mounted STARS system can bring two-fold benefits. On the one hand, with the aid of deep neural networks (DNNs), MADRL holds significant potential in processing echo signals for effective beam prediction at the RSU. On the other hand, the deployment of multi-agent enables independent control of target-mounted STARS from the vehicle, which can substantially reduce signaling overhead for information exchange in V2I communications.

Based on the analysis above, we propose a MADRL framework to achieve S\&C optimization for the sensing-assisted communication task. The main contributions of the paper are as follows:

\begin{itemize}

\item We present the system model and problem formulation for the considered target-mounted STARS-assisted ISAC system, where an RSU communicates with an in-vehicle user with the assistance of sensing signals reflected by the vehicle surface and the target-mounted STARS surface.
Our objective is to optimize the radar signal-to-noise ratio (SNR) and the achievable rate of the in-vehicle user by designing the transmit beamforming and receive filter of the RSU, as well as the reconfiguration of the STARS mounted on the target vehicle.
In order to solve this complicated and distributed design problem, the importance of adopting the MADRL framework is discussed and emphasized.

\item Next, we develop an MADRL framework to transform the optimization design problem into the Markov decision process (MDP). Based on the historical local-observable information, the RSU agent performs beam prediction while the Car agent configures target-mounted STARS. Compared to existing schemes, our end-to-end MADRL framework does not require complex transmission protocols for information sharing and additional measurements such as CSI.

\item Furthermore, to ensure efficient and stable learning for continuous decision-making, we design the multi-agent soft actor critic (MASAC) algorithm and the multi-agent proximal policy optimization (MAPPO) algorithm based on the proposed MADRL framework. A comprehensive analysis and comparison of two MADRL algorithms are provided to illustrate the superior performance compared to deterministic policies.


\item Finally, simulation results prove that MADRL algorithms can significantly enhance S\&C performance to realize sensing-assisted communication. Compared with STARS in the refraction-only mode, using STARS to reflect echo signals not only improves radar SNR but also assists communication to the in-vehicle user. The performance improvements in S\&C under various environmental conditions are also demonstrated.

\end{itemize}

The rest of this paper is organized as follows: Sec. \ref{section:A} introduces the target-mounted STARS-assisted ISAC system. After presenting a MADRL framework in Sec. \ref{section:C}, we propose an off-policy MASAC algorithm to optimize S\&C performance based on the MADRL framework in Sec. \ref{section:D}. To overcome the limitations of off-policy algorithms, we propose another on-policy MAPPO strategy in Sec. \ref{section:E}. Sec. \ref{section:F} provides simulation experiments and the analysis of different algorithms. Finally, we conclude our work in Sec. \ref{section:G}.

\section{System Model and Problem Formulation} \label{section:A}

\begin{figure}[t]
\centering
\includegraphics[width = 2.4 in]{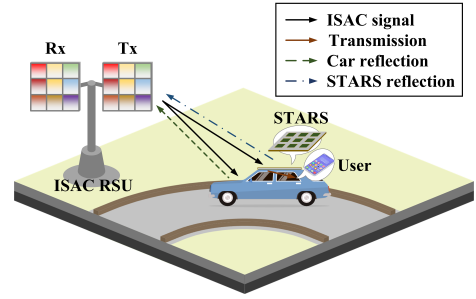}
\caption{The ISAC target-mounted STARS-assisted vehicular network.}
\label{fig:1}
\vspace{-0.3 cm}
\end{figure}

As shown in Fig. \ref{fig:1}, we consider a target-mounted STARS-assisted vehicular network, where one RSU communicates with an in-vehicle user with the assistance of STARS lodged on the vehicle surface. Recent research has demonstrated that the ISAC RSU can predict beams based on the echo signals reflected by the vehicles \cite{fliu2}-\cite{zhwang}, which avoids additional uplink pilot overhead. Meanwhile, we notice that the STARS can simultaneously reconfigure transmission and reflection links \cite{ywliu}. By employing the target-mounted STARS, on the one hand, the reflected echo signal can be further strengthened by appropriately adjusting the reflection coefficients of STARS to achieve more accurate beam prediction; on the other hand, by tuning the transmission coefficients of STARS, the transmission signal to the in-vehicle user can also be enhanced to overcome the high loss incurred when penetrating the vehicle. In our considered system, the uniform planar array (UPA) STARS has $M$ elements, and the set of its elements is represented as $\mathcal{M}=\{1,2,\ldots,M\}$.  We suppose that the RSU is equipped with $N_\mathrm{t}$ transmit antennas and $N_\mathrm{r}$ receive antennas, and the in-vehicle user is equipped with a single antenna. We divide the total ISAC service period $T$ into $N+1$ time slots, each of which has a $\Delta T$ duration. The CSI and motion parameters keep constant in the $n$-th time slot, where $\forall n \in \mathcal{N}=\{0,1,\ldots,N\}$ (i.e., $\Delta T = \frac{T}{N+1}$).

\vspace{-0.25 cm}
\subsection{Channel Model}
\begin{figure}[t]
\centering
\includegraphics[width = 2.4 in]{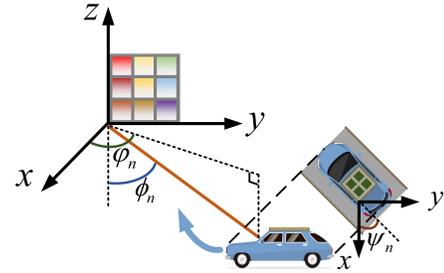}
\caption{The standard coordinate system in the vehicular network.}
\label{fig:2}
\vspace{-0.3 cm}
\end{figure}

As shown in Fig. \ref{fig:2}, we assume the transmit UPA and the receive UPA of the RSU are placed in the YOZ plane. Let $N_{\mathrm{t},y}$ and $N_{\mathrm{t},z}$ be the numbers of transmit antennas along the $y$-axis and $z$-axis, respectively. The elevation and azimuth angles of the RSU in the $n$-th time slot are denoted by $\varphi_n$ and $\phi_n$, respectively. Similarly, $N_{\mathrm{r},y}$ and $N_{\mathrm{r},z}$ represent the numbers of receive antennas along the $y$-axis and $z$-axis, respectively. Therefore, the steering vectors of RSU transmit antennas and receive antennas can be respectively expressed as
\begin{subequations}
\begin{align}
\mathbf{a}_{\mathrm{R}}(\varphi_n,\phi_n) & \triangleq \frac{1}{\sqrt{N_{\mathrm{t}}}}\Big[1,\ldots,e^{j\pi\big(n_1\mathrm{sin}(\varphi_n)\mathrm{sin}(\phi_n)+n_2\mathrm{cos}(\varphi_n)\big)}, \nonumber \\
&\ldots,e^{j\pi\big(N_{\mathrm{t},y}\mathrm{sin}(\varphi_n)\mathrm{sin}(\phi_n)+N_{\mathrm{t},z}\mathrm{cos}(\varphi_n)\big)}\Big]^{T},  \\
\mathbf{b}_{\mathrm{R}}(\varphi_n,\phi_n)& \triangleq \frac{1}{\sqrt{N_{\mathrm{r}}}}\Big[1,\ldots,e^{j\pi\big(n_3\mathrm{sin}(\varphi_n)\mathrm{sin}(\phi_n)+n_4\mathrm{cos}(\varphi_n)\big)}, \nonumber \\
&\ldots,e^{j\pi\big(N_{\mathrm{r},y}\mathrm{sin}(\varphi_n)\mathrm{sin}(\phi_n)+N_{\mathrm{r},z}\mathrm{cos}(\varphi_n)\big)}\Big]^{T}.
\end{align}
\end{subequations}

For the RSU-to-STARS path, the UPA steering vector of the STARS can be expressed as $\mathbf{a}_{\mathrm{S}}(\varphi_n^\ast,\phi_n^\ast)$, where $\varphi_n^\ast$ and $\phi_n^\ast$ represent the elevation and azimuth angles of the STARS in the $n$-th time slot with respect to its own coordinate system. We further model it in the standard coordinate system as shown in Fig. \ref{fig:2}, where $\psi_n$ denotes the angle between the vehicle and the $x$-axis in the $n$-th time slot. Thus, $\mathbf{a}_{\mathrm{S}}(\varphi_n^\ast,\phi_n^\ast)$ can be equivalently written as
\begin{equation}
\begin{aligned}[b]
\mathbf{a}_{\mathrm{S}}  \triangleq  &\frac{1}{\sqrt{M}}\Big[1,\ldots,e^{j\pi \mathrm{sin}(\overline{\psi}_n) \big(m_1\mathrm{sin}(\varphi_n)\mathrm{cos}(\phi_n)+m_2\mathrm{sin}(\varphi_n)\mathrm{sin}(\phi_n)\big)} \\
&,\ldots,e^{j\pi \mathrm{sin}(\overline{\psi}_n) \big(M_{s}\mathrm{sin}(\varphi_n)\mathrm{cos}(\phi_n)+M_{s}\mathrm{sin}(\varphi_n)\mathrm{sin}(\phi_n)\big)}\Big]^{T},
\end{aligned}
\end{equation}
where $\overline{\psi}_n = \mathrm{max}(\frac{\pi}{2}-\psi_n, \psi_n)$, and $M_s=\sqrt{M}$ represents the number of elements on each side of STARS.

For the STARS-to-user path, we assume the channel is quasi-static because it changes much more slowly compared to the RSU-STARS path. Thus the steering vector of STARS can be given as $\mathbf{a}_{\mathrm{S}}(\varphi_{\mathrm{0}}^\ast,\phi_{\mathrm{0}}^\ast)$. Therefore, the STARS-to-user channel $\mathbf{h}_{\mathrm{0}} \in \mathbb{C}^{M \times 1}$, the downlink RSU-to-STARS channel $\mathbf{G}^{\mathrm{DL}}_{n} \in \mathbb{C}^{M \times N_{\mathrm{t}}} $ and the uplink STARS-to-RSU channel $\mathbf{G}^{\mathrm{UL}}_{n} \in \mathbb{C}^{M \times N_{\mathrm{r}}}$ in the $n$-th time slot can be given as
\begin{subequations}
\begin{align}
\mathbf{G}^{\mathrm{DL}}_{n} &= \sqrt{MN_{\mathrm{t}}\alpha_{n}}  \mathbf{a}_{\mathrm{S}}(\varphi_n^\ast,\phi_n^\ast) \mathbf{a}_{\mathrm{R}}^{H}(\varphi_n,\phi_n), \label{2a}\\
\mathbf{G}^{\mathrm{UL}}_{n} &= \sqrt{MN_{\mathrm{r}}\alpha_{n}}  \mathbf{a}_{\mathrm{S}}(\varphi_n^\ast,\phi_n^\ast) \mathbf{b}_{\mathrm{R}}^{H}(\varphi_n,\phi_n), \label{2b}\\
\mathbf{h}_{\mathrm{0}} &= \sqrt{M\alpha_{\mathrm{0}}}  \mathbf{a}_{\mathrm{S}}(\varphi_{\mathrm{0}}^\ast,\phi_{\mathrm{0}}^\ast),\label{2c}
\end{align}
\end{subequations}
where the pass-loss coefficient $\alpha$ is modeled as $\alpha(d) = \alpha_{\mathrm{0}}(\frac{d}{d_{\mathrm{0}}})^{\zeta}$, $a_{\mathrm{0}}$ denotes the signal attenuation at the reference distance $d_{\mathrm{0}}$ and $\zeta$ is the pass loss exponent.

\subsection{Target-Mounted STARS Model}
STARS can simultaneously adjust the reflected signal and the transmitted signal by reconfiguring the EM property of each element. The transmission splitting ratio and the reflection splitting ratio of the $m$-th STARS element in the $n$-th time slot are written as $\beta_{n,m}^{\mathrm{T}}$ and $\beta_{n,m}^{\mathrm{R}}$, respectively, where $(\beta_{n,m}^{\mathrm{T}})^2 +(\beta_{n,m}^{\mathrm{R}})^2 = 1$ and $ \beta_{n,m}^{\mathrm{T}}, \beta_{n,m}^{\mathrm{R}} \in [0,1]$, $\forall n \in \mathcal{N}$, $\forall m \in \mathcal{M}$. Furthermore, the transmission phase-shift and reflection phase-shift of the $m$-th element in the $n$-th time slot are denoted as $\theta_{n,m}^{\mathrm{T}}$ and $\theta_{n,m}^{\mathrm{R}}$, respectively. We denote $\mathcal{F}$ and $B$ as the feasible phase-shift set and the phase-shift resolution to ensure $\theta_{n,m}^{\mathrm{T}}, \theta_{n,m}^{\mathrm{R}} \in \mathcal{F}$, where $\mathcal{F}\triangleq \{0, \frac{2\pi}{2^{B}},\ldots,\frac{2\pi\times(2^{B}-1)}{2^{B}}\}$. Accordingly, the matrices of the STARS transmission coefficients and reflection coefficients can be respectively expressed as
\begin{subequations}
\begin{align}
\mathbf{\Theta}_n^{\mathrm{T}}\triangleq\operatorname{diag}\big\{\beta_{n,1}^{\mathrm{T}} e^{j \theta_{n,1}^{\mathrm{T}}},~\beta_{n,2}^{\mathrm{T}} e^{j \theta_{n,2}^{\mathrm{T}}},\ldots, \beta_{n,M}^{\mathrm{T}} e^{j \theta_{n,M}^{\mathrm{T}}}\big\}, \\
\mathbf{\Theta}_n^{\mathrm{R}}\triangleq\operatorname{diag}\big\{\beta_{n,1}^{\mathrm{R}} e^{j \theta_{n,1}^{\mathrm{R}}},~\beta_{n,2}^{\mathrm{R}} e^{j \theta_{n,2}^{\mathrm{R}}},\ldots, \beta_{n,M}^{\mathrm{R}} e^{j \theta_{n,M}^{\mathrm{R}}}\big\}.
\end{align}
\end{subequations}

\vspace{-0.25 cm}
\subsection{Sensing Model}
We denote $s_{n}(t)$ as the ISAC signal transmitted by the RSU in the $n$-th time slot. As shown in Fig. \ref{fig:1}, the echo signal at the RSU consists of two components: One part is the signal reflected via the vehicle surface, and the other part is the signal reflected through the STARS surface using its reflection functionality. We denote transmit beamforming vector and receive beamforming/filtering in the $n$-th time slot as $\mathbf{w}_{n}^{\mathrm{t}} \in \mathbb{C}^{N_{\mathrm{t}} \times 1}$ and $\mathbf{w}_{n}^{\mathrm{r}} \in \mathbb{C}^{N_{\mathrm{r}} \times 1}$, respectively. The echo signal at the RSU is expressed as
\begin{equation}
\begin{aligned}
\bm{r}_n(t)= & e^{j2\pi\mu_{n}t}  \Big[ \underbrace{  G \beta \alpha_{n} \mathbf{b}_{\mathrm{R}}(\varphi_n,\phi_n) \mathbf{a}_{\mathrm{R}}^{H}(\varphi_n,\phi_n)  }_{\text {Car surface reflection }} \\
+ &\underbrace{(\mathbf{G}^{\mathrm{UL}}_{n})^{H}  \mathbf{\Theta}_n^{\mathrm{R}} \mathbf{G}^{\mathrm{DL}}_{n} }_{\text {STARS surface reflection}}  \Big] \mathbf{w}_{n}^{\mathrm{t}} s_{n}(t-\nu_{n})+\mathbf{z}_n(t),
\end{aligned}
\end{equation}
where $G=\sqrt{N_{\mathrm{t}}N_{\mathrm{r}}}$ is the total antenna array gain, $\beta\sim\mathcal{CN}(0, \sigma_{\mathrm{R}}^{2})$ represents the RCS of the vehicle surface, $\nu_{n}$ and $\mu_{n}$ are the time delay and Doppler frequency in the $n$-th time slot, respectively, which can be estimated by the matched-filtering method.

We use the matched filter to process the echo signal. Let $\Delta t$ denote the duration of one symbol and $\eta=\frac{\Delta T}{\Delta t}$ denote the total number of symbols processed by the matched filter in each time slot. Thus, the output signal after processing $\eta$ symbols by the matched filter in the $n$-th time slot can be given by \cite{fliu2}
\begin{equation}
\begin{aligned}\label{6}
\overline{\mathbf{r}}_n = & \sqrt{\eta}   \Big[ G \beta \alpha_{n}  \mathbf{b}_{\mathrm{R}}(\varphi_n,\phi_n) \mathbf{a}_{\mathrm{R}}^{H}(\varphi_n,\phi_n)   \\
+ &(\mathbf{G}^{\mathrm{UL}}_{n})^{H}  \mathbf{\Theta}_n^{\mathrm{R}} \mathbf{G}^{\mathrm{DL}}_{n}   \Big] \mathbf{w}_{n}^{\mathrm{t}} +\overline{\mathbf{z}}_n,
\end{aligned}
\end{equation}
where $\eta$ also denotes the matched-filtering gain, $ \overline{\mathbf{z}}_n\sim\mathcal{CN}(0, \sigma_{\mathrm{s}}^{2}\mathbf{I}_{N_{\mathrm{r}}})$ is the measurement noise. Next, we further process $\overline{\mathbf{r}}_n$ by using receive beamforming $\mathbf{w}_{n}^{\mathrm{r}}$, which is expressed as
\begin{equation}
\begin{aligned}\label{7}
(\mathbf{w}_{n}^{\mathrm{r}})^H \overline{\mathbf{r}}_n = & (\mathbf{w}_{n}^{\mathrm{r}})^H  \sqrt{\eta}   \Big[ G \beta \alpha_{n}  \mathbf{b}_{\mathrm{R}}(\varphi_n,\phi_n) \mathbf{a}_{\mathrm{R}}^{H}(\varphi_n,\phi_n)   \\
+ &(\mathbf{G}^{\mathrm{UL}}_{n})^{H}  \mathbf{\Theta}_n^{\mathrm{R}} \mathbf{G}^{\mathrm{DL}}_{n}   \Big] \mathbf{w}_{n}^{\mathrm{t}} + (\mathbf{w}_{n}^{\mathrm{r}})^H \overline{\mathbf{z}}_n.
\end{aligned}
\end{equation}

Therefore, the received SNR at the RSU in the $n$-th time slot is calculated as \cite{ktmeng}
\begin{equation}
\begin{aligned}\label{8}
\gamma_n^{\mathrm{RSU}} =  & \frac{\eta}{ (\mathbf{w}_{n}^{\mathrm{r}})^H \mathbf{w}_{n}^{\mathrm{r}} \sigma_{\mathrm{s}}^{2}}   \Big| (\mathbf{w}_{n}^{\mathrm{r}})^H  \big[ (\mathbf{G}^{\mathrm{UL}}_{n})^{H}  \mathbf{\Theta}_n^{\mathrm{R}} \mathbf{G}^{\mathrm{DL}}_{n}+ \\
&G \sigma_{\mathrm{R}} \alpha_{n}  \mathbf{b}_{\mathrm{R}}(\varphi_n,\phi_n) \mathbf{a}_{\mathrm{R}}^{H}(\varphi_n,\phi_n)\big] \mathbf{w}_{n}^{\mathrm{t}} \Big|^2.
\end{aligned}
\end{equation}

\vspace{-0.25 cm}
\subsection{Communication Model}
The target-mounted STARS is also employed to assist the wireless communication between the RSU and the in-vehicle user. We assume that the direct line-of-sight (LoS) path does not exist due to high penetration loss when passing through the vehicle. With the assistance of the STARS, the receive signal at the in-vehicle user is given by
\begin{equation}
\begin{aligned}
y_n(t)= & e^{j2\pi\mu_{n}t} \mathbf{h}_{\mathrm{0}}^{H}  \mathbf{\Theta}_n^{\mathrm{T}} \mathbf{G}^{\mathrm{DL}}_{n} \mathbf{w}_{n}^{\mathrm{t}} s_{n}(t)+\chi_n(t),
\end{aligned}
\end{equation}
where $ \chi_n(t)\sim\mathcal{CN}(0, \sigma_{\mathrm{c}}^{2})$ denotes the transmission noise. The received transmission SNR at the in-vehicle user and the achievable rate in $n$-th time slot are respectively given by
\begin{subequations}
\begin{align}
\gamma_n^{\mathrm{User}} &=  \frac{\big| \mathbf{h}_{\mathrm{0}}^{H}  \mathbf{\Theta}_n^{\mathrm{T}} \mathbf{G}^{\mathrm{DL}}_{n} \mathbf{w}_{n}^{\mathrm{t}} \big|^2 }{\sigma_{\mathrm{c}}^{2}}, \\
R_{n} &= \mathrm{log}_{2}(1+\gamma_n^{\mathrm{User}}).
\end{align}
\end{subequations}

We notice that the communication performance depends on the joint design of $\mathbf{w}_{n}^{\mathrm{t}}$ at the RSU and $\mathbf{\Theta}_n^{\mathrm{T}}$ at the target-mounted STARS. Besides, with the assistance of $\mathbf{w}_{n}^{\mathrm{r}}$ at the RSU and $\mathbf{\Theta}_n^{\mathrm{R}}$ at the target-mounted STARS, better sensing performance can provide more accurate beam prediction which can further enhance the communication performance.

\vspace{-0.25 cm}
\subsection{Problem Formulation}
We aim to jointly enhance S\&C performance to realize sensing-assisted communication with the assistance of the target-mounted STARS, where sensing performance is described by radar SNR $\gamma_n^{\mathrm{RSU}}$ and communication performance is measured by the achievable rate $R_{n}$. Since STARS should be directly controlled by the vehicle, the RSU and the vehicle maintain minimal information sharing to design beamforming and configure STARS, respectively. This multi-objective joint optimization problem is formulated as
\begin{subequations}
\begin{align}
\label{pa} &\max _{\substack{\mathbf{w}_{n}^{\mathrm{t}}, \mathbf{w}_{n}^{\mathrm{r}}, \mathbf{\Theta}_{n}^{\mathrm{R}}, \mathbf{\Theta}_{n}^{\mathrm{T}}}}~ (\gamma_n^{\mathrm{RSU}}, R_{n})\\ \label{pb}
&\quad~ \text { s.t. } ~~ \theta_{n,m}^{\mathrm{T}},~ \theta_{n,m}^{\mathrm{R}} \in \mathcal{F}, ~\forall n \in \mathcal{N},~ m \in \mathcal{M}, \\ \label{pc}
&\qquad\quad~~ \beta_{n,m}^{\mathrm{T}}, \beta_{n,m}^{\mathrm{R}} \in[0,1],~ \forall n \in \mathcal{N}, ~m \in \mathcal{M},\\ \label{pd}
&\qquad\quad~~ (\beta_{n,m}^{\mathrm{T}})^2 +(\beta_{n,m}^{\mathrm{R}})^2 = 1, ~\forall n \in \mathcal{N}, ~m \in \mathcal{M},\\ \label{pe}
&\qquad\quad~~ \| \mathbf{w}_{n}^{\mathrm{t}} \|^{2} \leq P, ~\forall n \in \mathcal{N},\\ \label{pf}
&\qquad\quad~~ \gamma_n^{\mathrm{RSU}} \geq \gamma_{\mathrm{min}}^{\mathrm{RSU}},~\forall n \in \mathcal{N},\\ \label{pg}
&\qquad\quad~~ \gamma_n^{\mathrm{User}} \geq \gamma_{\mathrm{min}}^{\mathrm{User}},~\forall n \in \mathcal{N},
\end{align}
\end{subequations}
where \eqref{pb}-\eqref{pd} are the phase-shift constraints and the energy splitting constraint of STARS, respectively. \eqref{pe} is the transmit power constraint and $P$ is the maximum transmit power at the RSU. $\gamma_{\mathrm{min}}^{\mathrm{RSU}}$ and $\gamma_{\mathrm{min}}^{\mathrm{User}}$ respectively denote the minimum received radar SNR and the minimum transmission SNR, which are used to ensure suitable S\&C performance during the ISAC service.

In this paper, we propose an end-to-end MADRL framework where two agents are respectively placed in the RSU and the vehicle (named as RSU/Car agent) for independent decision-making. Specifically, to minimize signaling overhead \cite{fliu2}, the RSU agent performs beam prediction while the Car agent pre-configures STARS to prepare for the ISAC service at the beginning of each time slot. In the following, we briefly explain the advantages of using MADRL framework compared with traditional algorithms in the target-mounted STARS ISAC system, as outlined below.

\begin{itemize}
\item \textbf{Less-observable real-time environmental information}: In dynamic vehicular networks, obtaining real-time environmental information (such as CSI) is extremely difficult for conducting effective beam prediction and STARS pre-configuration \cite{mnar}. Fortunately, MADRL is well-suited for tackling prediction problems based on the MDP framework \cite{ncl}, which allows agents to extract features from historical observable information through deep neural networks (DNNs).

\item \textbf{Incomplete and unshared status information}: To avoid frequent information sharing and significant pilot overhead, the RSU agent and the Car agent should perform beam prediction and STARS pre-configuration based on local environmental information, respectively. In MADRL approaches, each agent makes independent decisions by partial observation \cite{txli}.

\item \textbf{Multi-objective optimization}: We aim to improve S\&C performance to ultimately achieve better sensing-assisted communication for the in-vehicle user. However, trade-offs of multi-objectives in traditional optimization are static and rigid \cite{zhxiong}. To tackle this issue, MADRL allows for cooperation and competition among multiple agents \cite{txli}. Moreover, MADRL enables individual agents to make decisions for multiple objectives by designing its reward function.

\item \textbf{Continuous decision-making}: Each agent is required to maintain ISAC services throughout $N+1$ time slots. Unlike DL which relies on the extraction of features from vast static datasets, the fundamental nature of MADRL is updating policies dynamically through trial and error, whereas the MDP framework enables a continuous sequence of decision-making. Besides, MADRL can effectively tackle multi-step interaction problems by using lightweight networks.
\end{itemize}

Based on the above analysis, MADRL demonstrates tremendous potential in realizing sensing-assisted communication for the target-mounted STARS vehicular network.  In the next section, we will provide a generalized introduction to the proposed MADRL framework.

\section{MADRL Framework}\label{section:C}
In this section, we first present the multi-agent MDP structure for solving the considered problem (11). Then, we design the key elements of MDP while considering the challenges in the target-mounted STARS system. Finally, we propose the end-to-end MADRL framework.

\subsection{MDP Structure in Target-Mounted STARS Network}
The MDP structure provides a universal mathematical model for DRL and serves as a guiding principle for the design of most DRL algorithms. Assuming that the future state depends only on the previous state, the important parameters of DRL can be represented by tuple $(\mathcal{S},\mathcal{A},\mathcal{\pi},\mathcal{R}, \xi, \mathcal{P})$, in which $\mathcal{S}$ represents the \textbf{state} of the agent itself and the observed environment information. $\mathcal{A}$ and $\mathcal{\pi}$ represent the \textbf{action} taken based on the current state information and the \textbf{policy} for taking an action, respectively, which usually have connections in different DRL algorithms. $\mathcal{R}$ represents the \textbf{reward} based on the actions the agent has performed, which can be designed freely. The constant \textbf{discount factor} $\xi$ is used to balance the current reward with the future reward. Lastly, the agent proceeds to the next state $\mathcal{S^{\prime}}$ based on the \textbf{transition probabilities} $\mathcal{P}$ and repeats the aforementioned process.

We further introduce the multi-agent MDP structure in the target-mounted STARS system, where the RSU provides continuous ISAC services in total $N+1$ time slots for the in-vehicle user. From the DRL aspect, we consider the driving trajectory of the vehicle beginning from the starting point as one MADRL episode, with a maximum of $N+1$ DRL time steps. In the $n$-th time slot, the transitions of two agents are denoted as $T_n^{\mathrm{R}}=(s_{n-1}^{\mathrm{R}}, a_{n}^{\mathrm{R}}, r_{n}^{\mathrm{R}}, s_{n}^{\mathrm{R}}, d_{n})$ and $T_n^{\mathrm{C}} = (s_{n-1}^{\mathrm{C}}, a_{n}^{\mathrm{C}}, r_{n}^{\mathrm{C}}, s_{n}^{\mathrm{C}}, d_{n})$, respectively. As shown in Fig. \ref{fig:3}, at the beginning of the $n$-th time slot, both the RSU agent and the Car agent make decisions based on their respective observed \textbf{previous states} from the last time slot in the partially observable environment ($s_{n-1}^{\mathrm{R}}$ and $s_{n-1}^{\mathrm{C}}$), and generates corresponding \textbf{actions} ($a_{n}^{\mathrm{R}}$ and $a_{n}^{\mathrm{C}}$). Subsequently, the RSU communicates with the in-vehicle user and receives the echo signals reflected by the vehicle. At the same time, two agents interact with the real-time environment, observe the \textbf{current states} information ($s_{n}^{\mathrm{R}}$ and $s_{n}^{\mathrm{C}}$) and receive \textbf{rewards} ($r_{n}^{\mathrm{R}}$ and $r_{n}^{\mathrm{C}}$). Finally, agents utilize the observable information to determine whether the current episode is \textbf{done} ($d_{n}$) or proceed to the next time step. Throughout the iterative process, the agents store transitions for each time slot into the transitions buffer, which is used to periodically train agents to update better policies.

\begin{figure}[t]
\centering
\includegraphics[width = 3.0 in]{3.eps}
\caption{The MADRL framework for the target-mounted STARS-assisted vehicular network.}
\label{fig:3}
\vspace{-0.3 cm}
\end{figure}

\subsection{Configuration of Key Elements in MDP Structure}
Next, we provide detailed descriptions of each key transition element in the MDP structure in the target-mounted STARS system. It is important to emphasize that elements are designed using the partial environment information obtained by each agent. Meanwhile, in the high-speed and dynamic vehicular network, observing adequate environment information and obtaining accurate data such as CSI are challenging. The above issues undoubtedly pose challenges to our design. In order to maximize the acquisition of environmental information and achieve efficient learning, we design each key parameter as follows.

\begin{itemize}

\item \textbf{State}: The RSU agent can extract partial environment information by receiving signals processed by matched filtering. According to \eqref{6}, the state of the RSU agent in the $n$-th time slot $s_{n}^{\mathrm{R}}$ can be expressed as
\begin{equation}
\begin{aligned}\label{p11}
s_{n}^{\mathrm{R}} =\big \{ & \Re(\overline{\mathbf{r}}_n)^T,~ \Im(\overline{\mathbf{r}}_n)^T\big \},
\end{aligned}
\end{equation}
where $\Re(\cdot)$ and $\Im(\cdot)$ represent the real and imaginary parts, respectively, which integrate the state into the input format for the DNN. The dimension of $s_{n}^{\mathrm{R}}$ is $2N_{\mathrm{r}}$. For the state of the Car agent, considering the available environmental information at the vehicle end, $s_{n}^{\mathrm{C}}$ can be expressed as
\begin{equation}
\begin{aligned}\label{p12}
s_{n}^{\mathrm{C}} =\big \{ v_{n},~ \gamma_{n}^{\mathrm{User}} \big \},
\end{aligned}
\end{equation}
where $v_{n}$ represents the velocity of the vehicle in the $n$-th time slot. We ensure that $s_{n}^{\mathrm{C}}$ can be efficiently obtained by interacting with the local real-time environment.

\item \textbf{Action}: The output actions represent the decisions made by two agents. Specifically, the RSU agent designs the transmit beamforming and the receive filter while the Car agent configures the target-mounted STARS, which can be further described as follows:
\begin{subequations}\begin{align}
\label{p13}
a_{n}^{\mathrm{R}}&=\big \{\Re(\mathbf{w}_{n}^{\mathrm{t}})^T,\Re(\mathbf{w}_{n}^{\mathrm{r}})^T,
\Im(\mathbf{w}_{n}^{\mathrm{t}})^T,\Im(\mathbf{w}_{n}^{\mathrm{r}})^T\big\},\\
a_{n}^{\mathrm{C}}&=\!\big \{\theta_{n,1}^{\mathrm{R}},...,\theta_{n,m}^{\mathrm{R}},...\theta_{n,M}^{\mathrm{R}},
\theta_{n,1}^{\mathrm{T}},...,\theta_{n,m}^{\mathrm{T}},...\theta_{n,M}^{\mathrm{T}}, \nonumber\\
&\qquad\beta_{n,1}^{\mathrm{R}},...,\beta_{n,m}^{\mathrm{R}},...\beta_{n,M}^{\mathrm{R}}\big\}.\label{p14}
\end{align}\end{subequations}

The actions output by agents need further processing to satisfy the variable constraints of the considered problem. Specifically, the RSU agent processes $a_{n}^{\mathrm{R}}$ to satisfy constraint \eqref{pe}, while the Car agent processes $a_{n}^{\mathrm{C}}$ to satisfy constraints \eqref{pb} and \eqref{pc}. Meanwhile, the coefficient $\beta_{n,m}^{\mathrm{T}}$ is calculated according to constraint \eqref{pd}. The dimensions of the actions are $2(N_{\mathrm{t}}+N_{\mathrm{r}})$ and $3M$, respectively.

\item \textbf{Reward}: Similar to the state, the reward function must be designed using observable environmental information. For the RSU agent, we define the reward in the $n$-th time slot as
\begin{equation}\label{p15}
r_{n}^{\mathrm{R}} = f(\gamma_n^{\mathrm{Car}}) = 10\mathrm{lg}(\gamma_n^{\mathrm{Car}}).
\end{equation}

It is worth noting that the feedback from the Car agent via the uplink communication is not practical due to significant additional signaling overhead. Moreover, we design reward function $f(\cdot)$ to prevent DRL agents from becoming insensitive to small rewards, which may lead to negative learning guidance. At the same time, we define the reward function for the Car agent as
\begin{equation}\label{p16}
r_{n}^{\mathrm{C}} = \omega f(\gamma_{n}^{\mathrm{Car}})+ f(\gamma_{n}^{\mathrm{User}}),
\end{equation}
where $\omega$ is a weight factor the Car agent uses to balance S\&C performance more flexibly. For the Car agent, $\gamma_{n}^{\mathrm{User}}$ can be easily obtained by calculating the power of its received signals. 

\item \textbf{Done}: At any time slot, both the RSU agent and the Car agent must satisfy the minimum S\&C performance (i.e., constraints \eqref{pf} and \eqref{pg}). Moreover, if the ISAC service attains the maximum $N+1$ time slots, the ongoing episode comes to an end. Thus, we define an indicator $d_{n}$ to determine whether the episode terminates or not, as follows
\begin{equation}
\begin{aligned}\label{p17}
& \!\!\!\!\!\!\!\! d_{n} = ( \gamma_n^{\mathrm{RSU}} \geq \gamma_{\mathrm{min}}^{\mathrm{RSU}})  \ \cap \,  (\gamma_n^{\mathrm{User}} \geq \gamma_{\mathrm{min}}^{\mathrm{User}}) \, \cap \,  (n\leq N+1).
\end{aligned}
\end{equation}

\end{itemize}

\begin{algorithm}[t]
\renewcommand{\thealgorithm}{1}
\footnotesize
\caption{Proposed MADRL framework}
\label{alg:1}
\begin{algorithmic}[1]
\STATE{Initialize the RSU agent and the Car agent.}
\STATE{Initialize the transitions buffer.}
    \FOR {each episode $e$}\label{0-4}
        \FOR {each time slot $n=0,1,...,N$}\label{0-5}
         \STATE{For each agent $\mathrm{i}\in\{\mathrm{R},\mathrm{C}\}$, make action $a_{n}^{\mathrm{i}}$ \eqref{p13} and \eqref{p14} through the DNN network by inputting $s_{n-1}^{\mathrm{i}}$.}\label{0-5}
         \STATE{Finish the ISAC process in the $n$-th time slot.}\label{0-6}
         \STATE{For each agent $\mathrm{i}\in\{\mathrm{R},\mathrm{C}\}$, interact with the environment and observe $s_{n}^{\mathrm{i}}$ in \eqref{p11}, \eqref{p12} and $d_{n}$ in \eqref{p17}.}\label{0-7}
         \IF{Still train agents DNN network}\label{0-8}
            \STATE{For each agent $\mathrm{i}\in\{\mathrm{R},\mathrm{C}\}$, calculate $r_{n}^{\mathrm{i}}$ in \eqref{p15} and \eqref{p16}.}  \label{0-9}
            \STATE{For each agent $\mathrm{i}\in\{\mathrm{R},\mathrm{C}\}$, put $(s_{n-1}^{\mathrm{i}}, a_{n}^{\mathrm{i}}, r_{n}^{\mathrm{i}}, s_{n}^{\mathrm{i}}, d_{n})$ into transitions buffer and manage the transition \eqref{p18}.}\label{0-10}
            \STATE{\textbf{Jump to Algorithm \ref{alg:2} / \ref{alg:3}}}
            \IF{\emph{Meet certain conditions}}\label{0-11}
                \STATE{\emph{Use proposed MADRL algorithms for training.}}\label{0-12}
            \ENDIF\label{0-13}
            \STATE{\textbf{End Algorithm \ref{alg:2} / \ref{alg:3}}}
            \IF{$d_n$==1}\label{0-14}
                \STATE{Stop current training episode.}\label{0-15}
            \ENDIF\label{0-16}
         \ENDIF\label{0-17}
        \ENDFOR\label{0-18}
    \ENDFOR\label{0-19}
    \end{algorithmic}
\end{algorithm}

\subsection{Proposed MADRL Framework}
Based on the above analysis, as shown in Algorithm \ref{alg:1}, we propose the end-to-end MADRL framework where the S\&C optimization of the target-mounted STARS system can be converted into the MDP problem. We provide a detailed explanation of the proposed framework as follows.

We initiate a new episode when $n=0$ and each episode contains consecutive ISAC services with $N+1$ time slots. At the beginning of the $n$-th time steps, the RSU agent and the Car agent respectively predict corresponding beamforming $a_{n}^{\mathrm{R}}$ and STARS pre-configuration $a_{n}^{\mathrm{C}}$ based on the previous state $s_{n-1}^{\mathrm{i}}$ as step \ref{0-5}\footnote{It is important to note that we assume the partial environmental information observed at the starting point ($n=0$) is known for all agents, making $s_{-1}^{\mathrm{i}}$ readily accessible.}.  Then, two agents conduct ISAC service and observe respective local environmental information $s_{n}^{\mathrm{i}}$ as steps \ref{0-6}-\ref{0-7}. Steps \ref{0-8}-\ref{0-17} outline the iterative decision optimization process for agents. In step \ref{0-9}, each agent can obtain reward $r_{n}^{\mathrm{i}}$ based on $a_{n}^{\mathrm{i}}$ and $s_{n}^{\mathrm{i}}$ to evaluate current behavior\footnote{In our proposed framework, inter-agent downlink communication to calculate $r_{n}^{\mathrm{C}}$ \eqref{p16} is only required during the learning process, emphasizing that the fully-trained agents can independently make decisions without information exchange.}. In step \ref{0-10}, each agent stores the transition of the $n$-th time slot in the transitions buffer. For the buffer capable of accommodating $T_{\mathrm{max}}$ transitions, the transitions of each agent are managed and stored in sequential order, denoted as

\begin{equation}\label{p18}
T_n = (T_n^{\mathrm{R}},T_n^{\mathrm{C}})= (\mathbf{s}_{n-1}, \mathbf{a}_{n}, \mathbf{r}_{n}, \mathbf{s}_{n}, d_{n}).
\end{equation}

Steps \ref{0-11}-\ref{0-13} represent the training procedure to optimize policy for different MADRL algorithms. In the following sections, we propose two algorithms in order to ensure the efficiency of training from different perspectives.  Particularly, step \ref{0-11} is designed to ensure periodic training intervals to overcome the challenge of allocating adequate computational resources for training agents in every time slot.
In our proposed MADRL framework, the training interval is determined based on whether the algorithm is on-policy or off-policy.

In summary, we introduce a comprehensive MADRL framework for enhancing S\&C performance in target-mounted STARS-assisted vehicular networks. The framework ensures the seamless integration of various algorithms into our universal framework (in steps \ref{0-11}-\ref{0-14}). In the following sections, we present two highly efficient MADRL training algorithms designed to update the policy of each agent.

\section{Proposed off-policy MASAC Algorithm}\label{section:D}
In this section, we provide a comprehensive explanation of the training process for the off-policy MASAC algorithm. In off-policy strategies, a clear distinction arises between the target policy and the behavior policy, which allows agents to gain insights from an array of historical policies. Off-policy strategies exhibit distinctive characteristics by effectively utilizing all past transitions while simultaneously achieving a harmonious equilibrium between exploitation and exploration.
The MASAC algorithm can be considered a prime example of off-policy algorithms.  Compared with other off-policy deterministic algorithms, MASAC distinguishes itself by introducing the concept of maximum entropy, which facilitates exploration while simultaneously preventing the algorithm from converging to suboptimal solutions.  In the STARS-assisted vehicular network environment, the deployment of MASAC effectively reduces the overhead and training costs associated with agent-environment interactions. This benefit is accomplished by encouraging exploration and is more probable to offer substantial S\&C performance improvements.

\begin{figure}[t]
\centering
\includegraphics[width = 2.8 in]{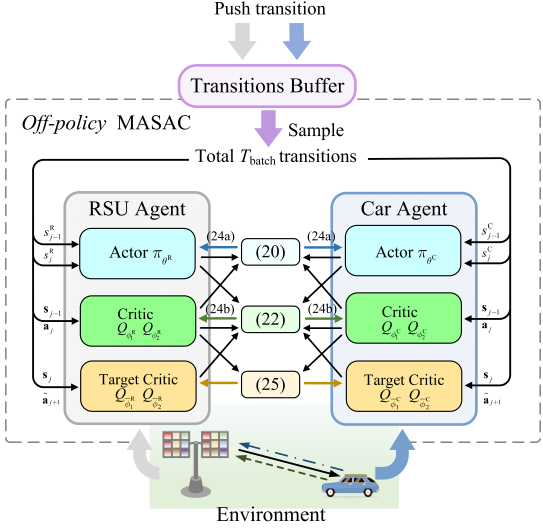}
\caption{Off-policy MASAC algorithm in target-mounted STARS-assisted vehicular network.}
\label{fig:4}
\vspace{-0.3 cm}
\end{figure}

We first introduce the architecture of MASAC. For each agent $\mathrm{i}\in\{\mathrm{R},\mathrm{C}\}$ in the MASAC algorithm, we employ the actor-critic (AC) framework, as illustrated in Fig. \ref{fig:4}. The AC networks are denoted as $\pi_{\theta^{\mathrm{i}}}(a_{j}^{\mathrm{i}}|s_{j-1}^{\mathrm{i}})$ and $Q_{\phi^{\mathrm{i}}_{l}}(\mathbf{s}_{j-1}, \mathbf{a}_{j})$, respectively, with $\theta^{\mathrm{i}}$ and $\phi^{\mathrm{i}}_{l}$ representing the DNN weight parameters. To tackle the over-estimation issue, the critic network employs two $Q$-functions (i.e., $l =1,2$), to provide more precise estimations of $Q$-values. Simultaneously, in order to further enhance the stability of learning, the target network $Q_{\overline{\phi}^{\mathrm{i}}_{l}}(\mathbf{s}_{j-1}, \mathbf{a}_{j})$ is adopted for the critic. We set the training interval to $E_{\mathrm{t}}$ episodes to alleviate deployment complexity (i.e., step \ref{0-12} of Algorithm \ref{alg:1}). The transitions buffer in the off-policy approach collects all transitions generated during the interaction process. Each agent randomly selects $T_{\mathrm{batch}}$ transitions from the transitions buffer for training.

Subsequently, we elaborate on the network update process during the training phase of the MASAC algorithm.
The primary goal of the SAC is to optimize policy entropy by identifying the highest possible reward. The basic framework of the SAC algorithm can be established through the following derived set of equations:
\begin{subequations}
\begin{align}\label{p19a}
V(S) &= \underset{\mathcal{A} \sim \pi}{\mathbb{E}}\Big[ Q(\mathcal{S},\mathcal{A})+\alpha H\big(\pi(\cdot|\mathcal{S})\big) \Big],\\ \label{p19b}
Q(\mathcal{S},\mathcal{A}) &= \underset{\mathcal{S}^{\prime} \sim \mathcal{P}}{\mathbb{E}}\big[ \mathcal{R} + \xi V(\mathcal{S}^{\prime})\big],\\ \label{p19c}
\pi &= \arg \min _{\pi} D_{\mathrm{KL}}\Big[ \pi (\mathcal{A}\! \mid \!\mathcal{S}) \|   \frac{\exp \big(\frac{1}{\alpha} Q (\mathcal{S}, \mathcal{A}) \big) }{Z(\mathcal{S})}   \Big],
\end{align}
\end{subequations}
where $H\big(\pi(\cdot|\mathcal{S})\big) = -\underset{\mathcal{A} \sim \pi}{\mathbb{E}}\log\pi(\mathcal{A}|\mathcal{S})$ denotes the entropy of policy $\pi$, and $\alpha$ is the temperature coefficient.
The expression for the soft value function, including the entropy term, is given by \eqref{p19a}.
The soft Bellman equation \eqref{p19b} comprehensively evaluates policies by combining entropy with the value function corresponding to the next state. Equation \eqref{p19c} minimizes the Kullback-Leibler (KL) divergence to seek a new updated policy that not only aims to achieve greater value but also maintains higher entropy. $Z(\mathcal{S})$ is the normalized distribution function.

Next, we introduce the MASAC algorithm in the target-mounted STARS-assisted vehicular network. For each agent, the update to critic networks is achieved by minimizing the soft Bellman residual. We define the set of $T_{\mathrm{batch}}$ managed transitions \eqref{p18} used for training as $\mathcal{D}$. The loss function for the critic can be written as
\begin{equation}\label{p20}
L(\phi^{\mathrm{i}}_{l}) = \underset{T_{j} \sim \mathcal{D}}{\mathbb{E}}\Big[ \big(Q_{\phi^{\mathrm{i}}_{l}}(\mathbf{s}_{j-1}, \mathbf{a}_{j}) - y^{\mathrm{i}} \big)^2 \Big],
\end{equation}
where $y^{\mathrm{i}}$ is the target soft value of each agent, which is denoted according to \eqref{p19a} and \eqref{p19b} as:
\begin{equation}\label{p21}
y^{\mathrm{i}} = r_{j}^{\mathrm{i}}+\xi \overline{d}_{j}\big( \min _{l=1,2} Q_{\overline{\phi}^{\mathrm{i}}_{l}}(\mathbf{s}_{j}, \mathbf{\widetilde{a}}_{j+1}) -
\alpha^{\mathrm{i}} \log \pi_{\theta^{\mathrm{i}}}(\widetilde{a}_{j+1}^{\mathrm{i}}|s_{j}^{\mathrm{i}}) \big),
\end{equation}
where $\overline{d}_{j} = 1-d_{j}$. It is important to emphasize that $\widetilde{a}_{j}^{\mathrm{i}}$ is not $a_{j}^{\mathrm{i}}$ stored in transitions, but rather an action sampled from the probability distribution of the policy based on the output of the actor network, denoted as $\widetilde{a}_{j}^{\mathrm{i}}\sim \pi(\cdot|{s}_{j-1})$. From equation \eqref{p19c}, we can derive the loss function for each actor as:
\begin{equation}\label{p22}
L(\theta^{\mathrm{i}}) = \underset{T_{j} \sim \mathcal{D}}{\mathbb{E}} \big[ \alpha^{\mathrm{i}} \log \pi_{\theta^{\mathrm{i}}}(\widetilde{a}_{j}^{\mathrm{i}}|s_{j-1}^{\mathrm{i}})
-\min _{l=1,2} Q_{\overline{\phi}^{\mathrm{i}}_{l}}(\mathbf{s}_{j-1}, \mathbf{\widetilde{a}}_{j}) \big].
\end{equation}

In order to balance exploration and exploitation, we employ dynamic entropy adjustment to adapt to different learning steps and the target entropy is expressed as $ \widehat{H}^{\mathrm{i}}= \mathrm{dim}(a_{j}^{\mathrm{i}})$. Therefore, the corresponding loss function is given by
\begin{equation}\label{p23}
L(\alpha^{\mathrm{i}}) = \underset{T_{j} \sim \mathcal{D}}{\mathbb{E}} \big[ \alpha^{\mathrm{i}} \log \pi_{\theta^{\mathrm{i}}}(\widetilde{a}_{j}^{\mathrm{i}}|s_{j-1}^{\mathrm{i}})
- \alpha^{\mathrm{i}} \widehat{H}^{\mathrm{i}}\big].
\end{equation}

To achieve better policy for each agent, we update the parameters of AC networks and the entropy through gradient descent using the following operations:
\begin{subequations}
\begin{align}\label{p24a}
\phi^{\mathrm{i}}_{l} &\longleftarrow \phi^{\mathrm{i}} -\beta_{\phi^{\mathrm{i}}_{l}} \cdot \nabla_{\phi^{\mathrm{i}}_{l}} L(\phi^{\mathrm{i}}_{l}),\\\label{p24b}
\theta^{\mathrm{i}} &\longleftarrow \theta^{\mathrm{i}} -\beta_{\theta^{\mathrm{i}}} \cdot \nabla_{\theta^{\mathrm{i}}} L(\theta^{\mathrm{i}}),\\\label{p24c}
\alpha^{\mathrm{i}} &\longleftarrow \alpha^{\mathrm{i}} -\beta_{\alpha^{\mathrm{i}}} \cdot \nabla_{\alpha^{\mathrm{i}}} L(\alpha^{\mathrm{i}}).
\end{align}
\end{subequations}

\begin{algorithm}[t]
\renewcommand{\thealgorithm}{2}
\footnotesize
\caption{{Proposed off-policy MASAC algorithm}}
\label{alg:2}
\begin{algorithmic}[1]
\STATE{\textbf{Start from step \ref{0-11} of Algorithm \ref{alg:1}.}}
\IF{$e \% E_{\mathrm{t}} ==0$}
    \STATE{Random sample transitions from the transitions buffer.}
    \STATE{For each agent $\mathrm{i}\in\{\mathrm{R},\mathrm{C}\}$, compute target values in \eqref{p21}.}
    \STATE{For each agent $\mathrm{i}\in\{\mathrm{R},\mathrm{C}\}$, update Q-functions by \eqref{p20} and \eqref{p24a}.}
    \STATE{For each agent $\mathrm{i}\in\{\mathrm{R},\mathrm{C}\}$, update policy by \eqref{p22} and \eqref{p24b}.}
    \STATE{For each agent $\mathrm{i}\in\{\mathrm{R},\mathrm{C}\}$, adjust temperature in \eqref{p23} and \eqref{p24c}.}
    \STATE{For each agent $\mathrm{i}\in\{\mathrm{R},\mathrm{C}\}$, soft update target networks \eqref{p25}.}
\ENDIF
\STATE{\textbf{Back to step \ref{0-13} of Algorithm \ref{alg:1}.}}
\end{algorithmic}
\end{algorithm}

Finally, soft updates are employed to update the parameters of the target critic, mitigating abrupt parameter changes and ensuring a smoother learning process:
\begin{equation}\label{p25}
\overline{\phi}^{\mathrm{i}}_{l} = \tau_{\overline{\phi}^{\mathrm{i}}_{l}} \overline{\phi}^{\mathrm{i}}_{l} + (1 - \tau_{\overline{\phi}^{\mathrm{i}}_{l}}) \phi^{\mathrm{i}}_{l}.
\end{equation}

Algorithm \ref{alg:2} summarizes the off-policy MASAC training process, which is built upon the MADRL framework presented in Sec. \ref{section:C}.
The findings from the simulation results provide evidence that the MASAC algorithm presents notable benefits as compared to other deterministic strategies. Nevertheless, it is also important to acknowledge that off-policy strategies exhibit certain limitations due to their training methodologies \cite{ncl}.
The off-policy algorithms utilize all past transitions to ensure comprehensive learning. However, it will lead to decreased learning efficiency and slow convergence. Besides, the learning process of the off-policy strategy is unstable as it may exploit poor transitions. Furthermore, the practical implementation of off-policy algorithms is hindered by their sensitivity to parameters.
Considering these potential drawbacks of applying the MASAC algorithm in the target-mounted STARS vehicular network, we present an on-policy approach in the next section.

\section{Proposed on-policy MAPPO Algorithm}\label{section:E}
\begin{figure}[t]
\centering
\includegraphics[width = 2.8 in]{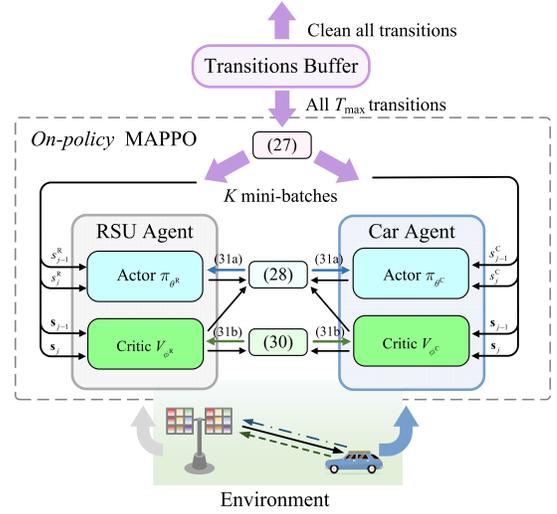}
\caption{On-policy MAPPO algorithm in target-mounted STARS-assisted vehicular network.}
\label{fig:5}
\end{figure}

In this section, we propose to deploy an on-policy MAPPO algorithm, which can maintain both effective exploration and stable learning capabilities, in the MADRL framework introduced in Sec. \ref{section:C}.
The MAPPO algorithm successfully tackles the convergence difficulties encountered by traditional policy gradient (PG) algorithms. In our environment setup, as illustrated in Fig. \ref{fig:5}, the transitions buffer stores $T_{\mathrm{max}}$ transitions generated from the interactions between the current policy and the environment, and all of these transitions are utilized for learning purposes. To mitigate gradient variance and enhance stability, we adopt a strategy of dividing all transitions $\mathcal{D}$ into $\mathcal{D}_K$ distinct mini-batches during each learning process. Once the learning process is complete, the transitions buffer is cleared to prepare for storing transitions from the new policy. Moreover, to ensure an accurate estimation of the latest policy, target networks are not employed. For each agent, the AC networks are represented by $\pi_{\theta^{\mathrm{i}}}(a_{j}^{\mathrm{i}}|s_{j-1}^{\mathrm{i}})$ and $V_{\phi^{\mathrm{i}}}(\mathbf{s}_{j-1})$, respectively.

\begin{algorithm}[t]
\renewcommand{\thealgorithm}{3}
\footnotesize
\caption{{Proposed on-policy MAPPO algorithm}}
\label{alg:3}
\begin{algorithmic}[1]
\STATE{\textbf{Start from step \ref{0-11} of Algorithm \ref{alg:1}.}}
\IF{ reach $ T_{\mathrm{max}}$ transitions}
    \STATE{For each agent $\mathrm{i}\in\{\mathrm{R},\mathrm{C}\}$, compute advantage estimates in \eqref{p27}.}
    \STATE{Randomly partition $T_{\mathrm{max}}$ transitions into $K$ mini-batches.}
    \FOR {each mini-batch $k=0,1,...,K$}
        \STATE{For each agent $\mathrm{i}\in\{\mathrm{R},\mathrm{C}\}$, update the policy by \eqref{p28} and \eqref{p31a}.}
        \STATE{For each agent $\mathrm{i}\in\{\mathrm{R},\mathrm{C}\}$, fit value function by \eqref{p30} and \eqref{p31b}.}
    \ENDFOR
    \STATE{Clear all transitions in the transitions buffer.}
\ENDIF
\STATE{\textbf{Back to step \ref{0-13} of Algorithm \ref{alg:1}.}}
\end{algorithmic}
\end{algorithm}

\begin{figure*}[t]
 \begin{minipage}{0.4\linewidth}
 \centering
 \includegraphics[width=2.5in]{6.eps}
 \caption{The illustration of considered target-mounted STARS-assisted vehicular network.}
 \label{fig:6}
 \end{minipage}%
 \begin{minipage}{0.68\linewidth}
 \footnotesize
	\centering
    \makeatletter\def\@captype{table}\makeatother
    \caption{Environment parameters and MADRL hyperparameters}
    \label{table:1}
    \renewcommand\arraystretch{1}{
	\begin{tabular}{|c|c|c|c|c|c|}
    \hline
    Environment  & \multirow{2}*{Values} & MASAC  & \multirow{2}*{Values} & MAPPO & \multirow{2}*{Values}\\
    parameters &  & parameters &  & parameters &  \\
    \hline
	$M$   &	36 &  $\omega$ & 1 &  $\omega$ & 1 \\
    \hline
    $N_{\mathrm{t}},N_{\mathrm{r}}$  &  25 & $\xi$  &0.99 &  $\xi$  &0.99 \\
    \hline
    $P$   &   30dBm &  $T_{\mathrm{batch}}$ & 256 &  $\epsilon$ & 0.2 \\
    \hline
    $\Delta T$   &   0.1s & $E_{\mathrm{t}}$ & 5 & $\lambda$ & 0.95 \\
    \hline
    $N$   &   10  & $\alpha^{\mathrm{i}}$ & 0.1 & $T_{\mathrm{max}}$ & 256 \\
    \hline
    $\eta$   &   10 &  $\beta_{\phi^{\mathrm{i}}_{l}}$ & $ 5\times 10^{-5}$ & $K$ & 64 \\
    \hline
    $B$   &   3  & $ \beta_{\theta^{\mathrm{i}}} $ & $5\times 10^{-5}$ & $\beta_{\phi^{\mathrm{i}}}$ & $2\times 10^{-4}$\\
    \hline
    $\sigma_{\mathrm{s}}^{2}, \sigma_{\mathrm{c}}^{2}$   &   -80dBm &  $\tau_{\overline{\phi}^{\mathrm{i}}_{l}}$ &    $5\times 10^{-2}$   & $\beta_{\theta^{\mathrm{i}}}$ & $2\times 10^{-4}$\\
    \hline
	\end{tabular}}
 \end{minipage}
 \end{figure*}

Subsequently, we provide a detailed explanation of implementing the MAPPO algorithm in the target-mounted STARS-assisted vehicular network. MAPPO is a policy optimization algorithm. Specifically, the old policy and the new policy are denoted as $\pi_{\theta^{\mathrm{i}}}^{\mathrm{o}}$ and $\pi_{\theta^{\mathrm{i}}}^{\mathrm{n}}$, respectively. The theoretical objective can be expressed as:
\begin{subequations}
\begin{align}
\label{p26a} & \mathop{\max}\limits_{\theta^{\mathrm{i}}}~ \underset{T_{j} \sim \mathcal{D}_k}{\mathbb{E}} \big[  \frac{\pi_{\theta^{\mathrm{i}}}^{\mathrm{n}}(\widetilde{a}_{j}^{\mathrm{i}} \mid s_{j-1}^{\mathrm{i}})}{\pi_{\theta^{\mathrm{i}}}^{\mathrm{o}}(a_{j}^{\mathrm{i}} \mid s_{j-1}^{\mathrm{i}})} A(a_{j}^{\mathrm{i}},s_{j-1}^{\mathrm{i}}) \big], \\ \label{p26b}
&~\text { s.t. } ~~ \underset{T_{j} \sim \mathcal{D}_k}{\mathbb{E}} \big[  \frac{\pi_{\theta^{\mathrm{i}}}^{\mathrm{n}}(\widetilde{a}_{j}^{\mathrm{i}} \mid s_{j-1}^{\mathrm{i}})}{\pi_{\theta^{\mathrm{i}}}^{\mathrm{o}}(a_{j}^{\mathrm{i}} \mid s_{j-1}^{\mathrm{i}})} -1 \big]\leq\epsilon,~\forall k,~ \forall j, ~\forall \mathrm{i },
\end{align}
\end{subequations}
where constraint \eqref{p26b} serves the purpose of limiting the disparity between the new and old policies to a reasonable extent, while also ensuring that the new policy remains feasible. In our MAPPO algorithm, we employ generalized advantage estimation (GAE) to optimize the advantage function, effectively managing the trade-off between variance and bias.  Additionally, we estimate the advantage function using the temporal difference (TD)-target, which is expressed as follows:
\begin{equation}
\begin{aligned}\label{p27}
A(a_{j}^{\mathrm{i}},s_{j-1}^{\mathrm{i}})&= \delta_{j}^{\mathrm{i}} +\sum_{l=1}^{T_{\mathrm{max}}}(\xi \lambda)^l \overline{d}_{j} \delta_{j+l}^{\mathrm{i}},  \\
&=\delta_{j}^{\mathrm{i}}+\xi \lambda  \overline{d}_{j} A(a_{j+1}^{\mathrm{i}},s_{j}),
\end{aligned}
\end{equation}
where $\delta_{j}^{\mathrm{i}} = r_{j}^{\mathrm{i}} + \xi \overline{d}_{j} V_{\phi^{\mathrm{i}}}(\mathbf{s}_{j})- V_{\phi^{\mathrm{i}}}(\mathbf{s}_{j-1})$ is the advantage function estimated by critic value network, and $\lambda$ denotes the GAE smooth factor. Based on \eqref{p26a} and \eqref{p26b}, we utilize PPO-Clip to construct the loss function for the actor network, given by
\begin{equation}
\begin{aligned}\label{p28}
&L(\theta^{\mathrm{i}}) = \\
&\underset{T_{j} \sim \mathcal{D}_k}{\mathbb{E}} \bigg [ \min \Big(\frac{\pi_{\theta^{\mathrm{i}}}^{\mathrm{n}}(\widetilde{a}_{j}^{\mathrm{i}} \mid s_{j-1}^{\mathrm{i}})}{\pi_{\theta^{\mathrm{i}}}^{\mathrm{o}}(a_{j}^{\mathrm{i}} \mid s_{j-1}^{\mathrm{i}})}
 A(a_{j}^{\mathrm{i}},s_{j-1}^{\mathrm{i}}), g\big ( A(a_{j}^{\mathrm{i}},s_{j-1}^{\mathrm{i}})\big) \Big) \bigg],
 \end{aligned}
\end{equation}
where $g(\cdot)$ is used to satisfy the constraint \eqref{p26b}, which is given by

\begin{equation}\label{p29}
g(A)= \begin{cases}(1+\epsilon) A , &  \mathrm{if} \ A \geq 0, \\ (1-\epsilon) A , & \mathrm{if} \ A<0.\end{cases}
\end{equation}

\nid The loss function of the critic network can be expressed as
\begin{equation}\label{p30}
L(\phi^{\mathrm{i}}) = \underset{T_{j} \sim \mathcal{D}_{k}}{\mathbb{E}}\Big[ r_{j}^{\mathrm{i}} + \xi \overline{d}_{j} V_{\phi^{\mathrm{i}}}(\mathbf{s}_{j})- V_{\phi^{\mathrm{i}}}(\mathbf{s}_{j-1}) \Big].
\end{equation}

Finally, we replace actor parameters to optimize policies by using gradient ascent and update the critic network to evaluate policies reasonably by using gradient descent:
\begin{subequations}
\begin{align}\label{p31a}
\theta^{\mathrm{i}} &\longleftarrow \theta^{\mathrm{i}} +\beta_{\theta^{\mathrm{i}}} \cdot \nabla_{\theta^{\mathrm{i}}} L(\theta^{\mathrm{i}}),\\ \label{p31b}
\phi^{\mathrm{i}} &\longleftarrow \phi^{\mathrm{i}} -\beta_{\phi^{\mathrm{i}}} \cdot \nabla_{\phi^{\mathrm{i}}} L(\phi^{\mathrm{i}}).
\end{align}
\end{subequations}

Algorithm \ref{alg:3} outlines the training process of MAPPO. From the perspective of DNN structures, it is clear that the MAPPO agents are more lightweight without critic networks compared to the on-policy MASAC algorithm. Meanwhile, MAPPO realizes a significant reduction in the storage requirements for the transitions buffer, which only needs to save transitions generated by the current policy. However, on-policy algorithms have limitations, such as converging to the local optima. In the next section, we present simulations to compare the performance of two proposed MADRL algorithms in the target-mounted STARS-assisted vehicular network.

\section{Simulation Results}\label{section:F}
In this section, we provide extensive simulations to comprehensively evaluate the performance of MASAC and MAPPO algorithms within the proposed MADRL framework. As depicted in Fig. \ref{fig:6}, we consider a target-mounted STARS-assisted vehicular network, where the RSU is located at (-50m, 0, 30m) and serves the vehicle moving along the curve road with varying speeds. The deployment altitude of the STARS is set at 2m.
The path-loss exponents of the RSU-STARS and STARS-user channels are set to 2.5 and 2.8, respectively. The environmental information at the start point (i.e., $n=0$ at (10m,50m,0)) is known. For each agent, $s^{\mathrm{i}}_{-1}$ is obtained under the conditions of $\theta_{-1,m}^{\mathrm{R}}=\theta_{-1,m}^{\mathrm{T}}=0$ and $(\beta_{-1,m}^{\mathrm{R}})^2=(\beta_{-1,m}^{\mathrm{T}})^2=0.5$. The minimum SNR to ensure satisfactory S\&C performance is set as $\gamma_{\mathrm{min}}^{\mathrm{RSU}}=\gamma_{\mathrm{min}}^{\mathrm{User}}=10\mathrm{dB}$. Other default environment parameters are listed in Table \ref{table:1}.

\begin{figure}[t]
\centering
\includegraphics[width = 3.3 in]{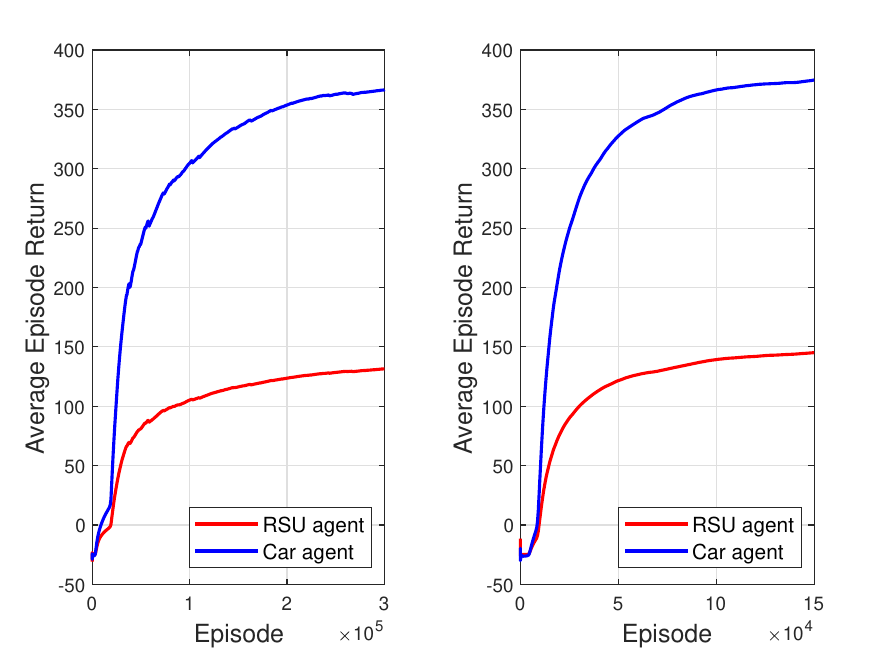}
\footnotesize \hbox{(a) Off-policy MASAC Algorithm. \hspace{0.1cm} (b) On-policy MAPPO Algorithm.}
\caption{Return convergence curves. }
\label{f1}
\vspace{-0.3 cm}
\end{figure}

We first verify the convergence performance of MASAC and MAPPO algorithms within the proposed MADRL framework. Here, we denote $G^{\mathrm{i}}=\sum_{n=0}^{N}r_{n}^{\mathrm{i}}$ as the cumulative reward (total return) accumulated by each agent during interactions with the environment. In an effort to clearly evaluate convergence performance, we calculate the average return over all episodes. For the $e$-th episode, the average episode return is denoted as $\overline{G}^{\mathrm{i}}_{e} = \frac{1}{e}\sum_{e}G^{\mathrm{i}}$. Fig. \ref{f1} illustrates the convergence performance of two algorithms under default environmental parameters, with the hyperparameters for each algorithm listed in Table \ref{table:1}. It can be observed that both algorithms achieve stable convergence performance within our proposed MADRL framework. Additionally, the proposed MADRL framework excels in learning under limited environmental information by effectively extracting meaningful features from the local-observable environment.

\begin{figure}[t]
\centering
\includegraphics[width = 3.3 in]{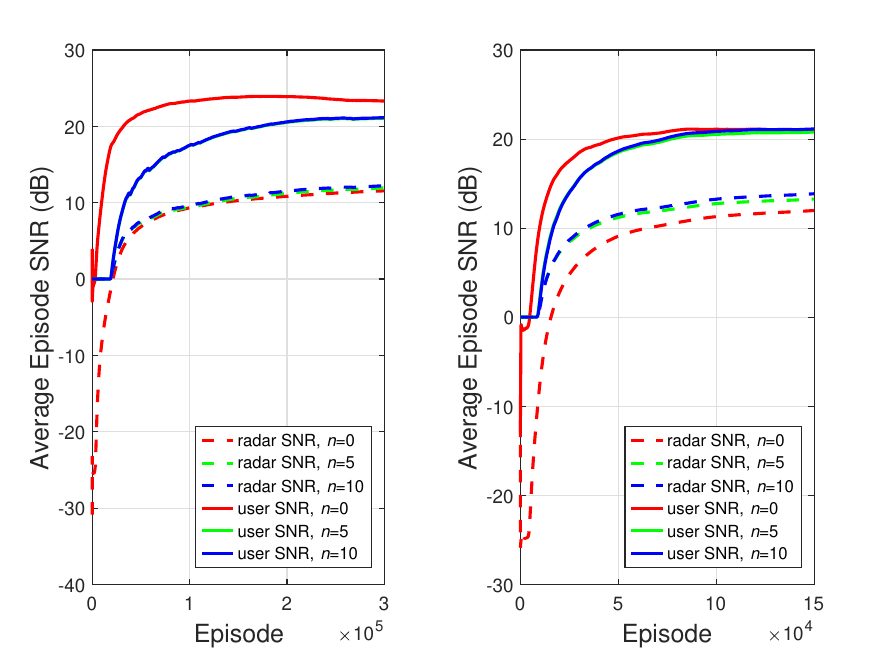}
\footnotesize \hbox{(a) Off-policy MASAC Algorithm. \hspace{0.1cm} (b) On-policy MAPPO Algorithm.}
\caption{SNR convergence curves. }
\label{f2}
\vspace{-0.3 cm}
\end{figure}

We delve further into analyzing the convergence of S\&C performance in each slot, focusing on three selected time slots $(n=0, 5, 10)$. We utilize average episode SNR $\overline{\gamma_n}^{\mathrm{RSU}} = \frac{1}{e}\sum_{e}\gamma_n^{\mathrm{RSU}}$ and $\overline{\gamma_n}^{\mathrm{User}} = \frac{1}{e}\sum_{e}\gamma_n^{\mathrm{RSU}}$, as metrics to evaluate the convergence of S\&C performance, respectively. Fig. \ref{f2} presents the convergence results, demonstrating that the radar SNR for the RSU and the communication SNR for the in-vehicle user are significantly improved and can exceed $\gamma_{\mathrm{min}}^{\mathrm{RSU}}$ and $\gamma_{\mathrm{min}}^{\mathrm{User}}$ through consistently learning. Both agents exhibit robust convergence performance in complex and unknown environments ($n>0$), proving that MASAC and MAPPO algorithms are effective in S\&C optimization.

\begin{figure}[t]
\centering
\includegraphics[width = 3.3 in]{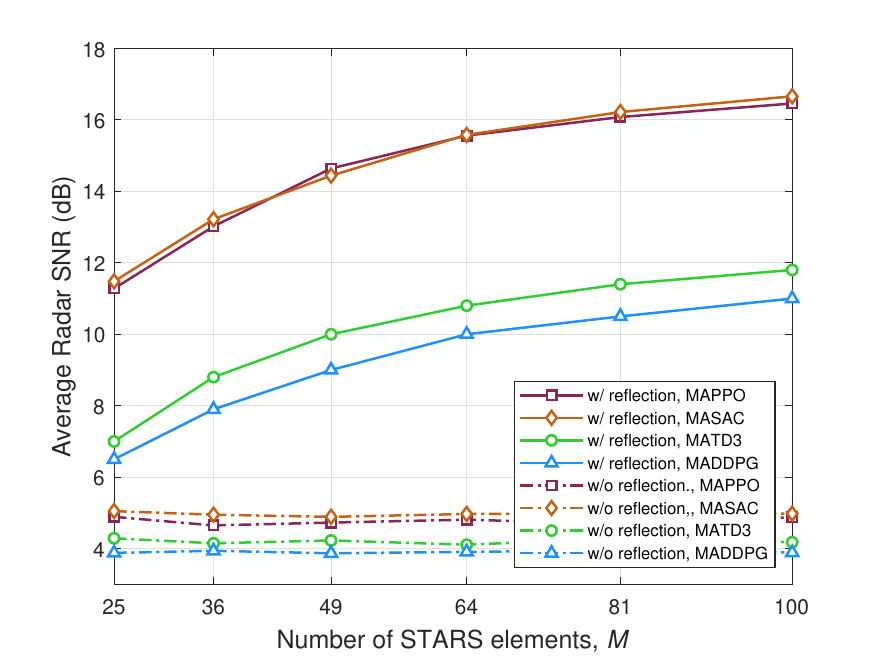}
\caption{Sensing performance versus the number of STARS elements $M$.}
\label{f3a}
\vspace{-0.3 cm}
\end{figure}

\begin{figure}[t]
\centering
\includegraphics[width = 3.3 in]{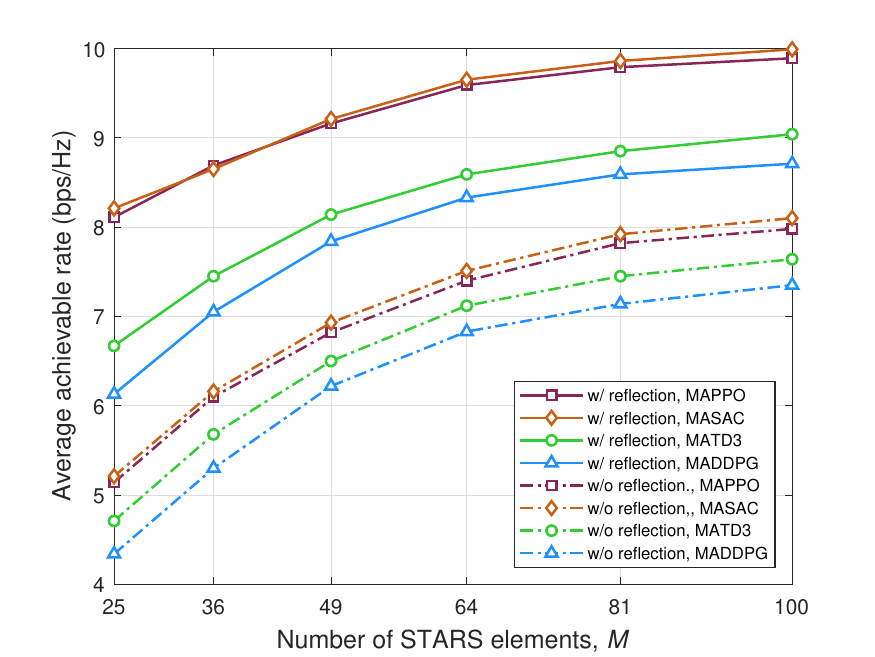}
\caption{Communication performance versus the number of STARS elements $M$.}
\label{f3b}
\vspace{-0.3 cm}
\end{figure}

\begin{figure}[t]
\centering
\includegraphics[width = 3.3 in]{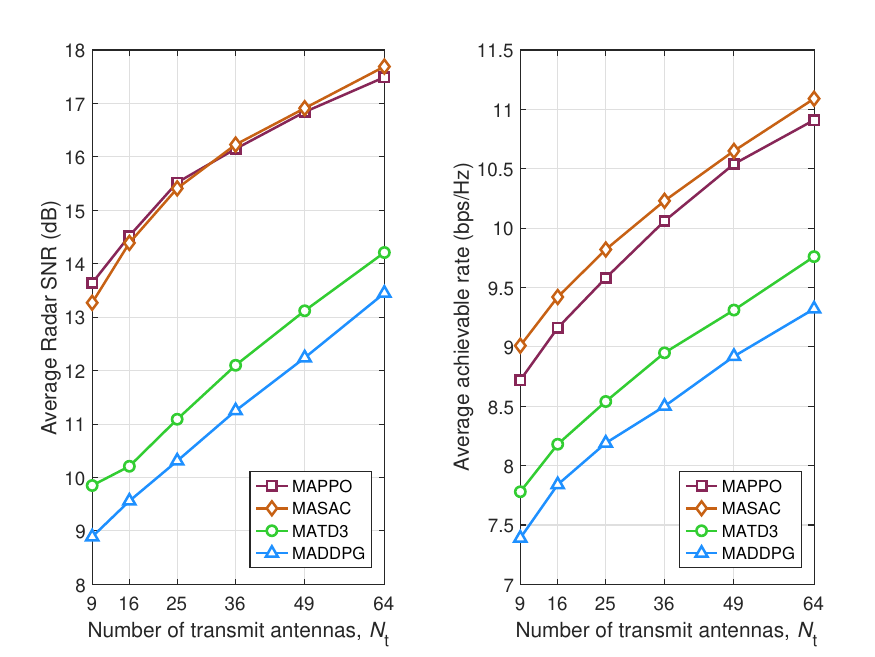}
\footnotesize \hbox{\hspace{0.5cm}(a) Sensing performance. \hspace{0.2cm} (b) Communication performance.}
\caption{Sensing and communication performance versus the number of transmit antennas $N_{\mathrm{t}}$ ($M=64$).}
\label{f4}
\vspace{-0.3 cm}
\end{figure}

\begin{figure}[t]
\centering
\includegraphics[width = 3.3 in]{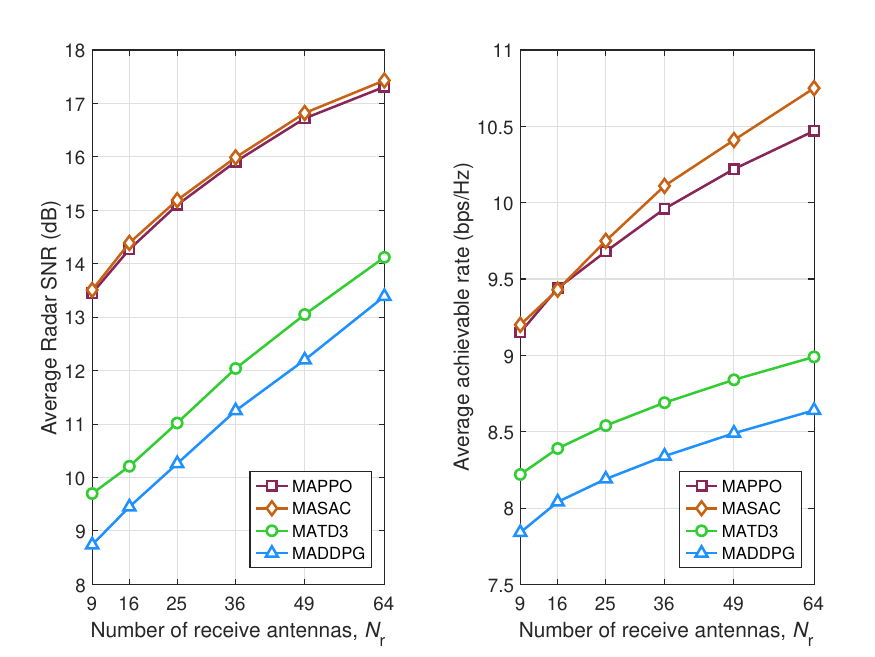}
\footnotesize \hbox{\hspace{0.5cm}(a) Sensing performance. \hspace{0.2cm} (b) Communication performance.}
\caption{Sensing and communication performance versus the number of receive antennas $N_{\mathrm{r}}$ ($M=64$).}
\label{f5}
\vspace{-0.3 cm}
\end{figure}

\begin{figure}[t]
\centering
\includegraphics[width = 3.3 in]{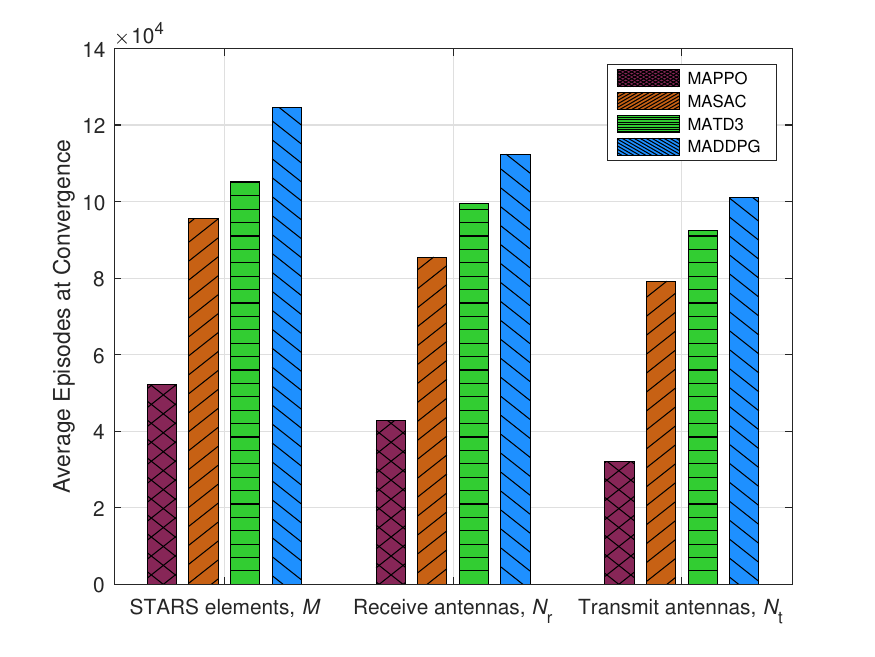}
\caption{Average convergence episodes under different cases ($M, N_{\mathrm{t}}$ and $N_{\mathrm{r}}$).}
\label{f6}
\vspace{-0.3 cm}
\end{figure}

\begin{figure}[t]
\centering
\includegraphics[width = 3.3 in]{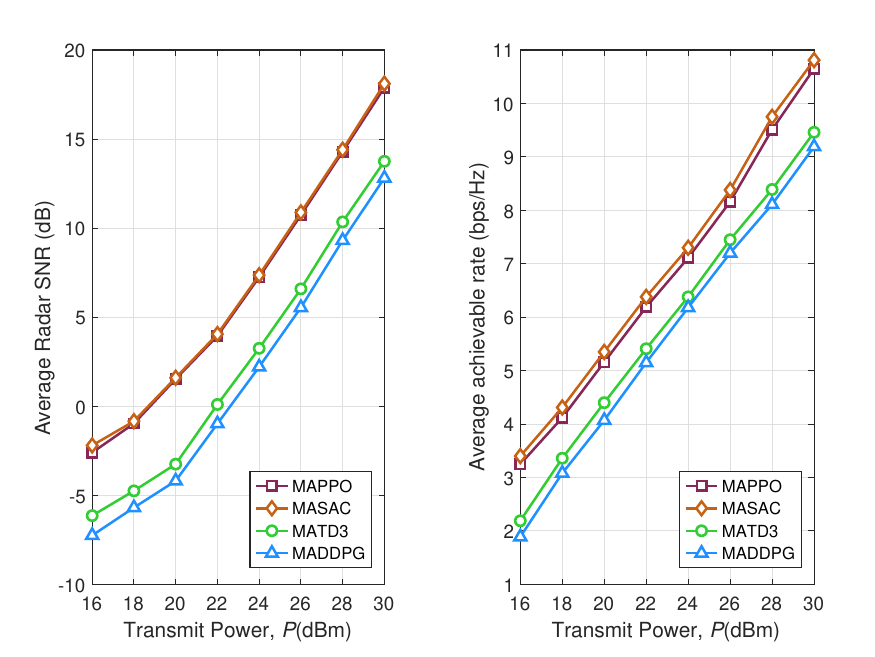}
\footnotesize \hbox{\hspace{0.5cm}(a) Sensing performance. \hspace{0.2cm} (b) Communication performance.}
\caption{Sensing and communication performance versus transmit power $P$ ($M=64$, $N_{\mathrm{t}}=N_{\mathrm{r}}=36$).}
\label{f7}
\vspace{-0.3 cm}
\end{figure}

Next, we conduct simulations to further evaluate the S\&C performance based on the converged agent models in different scenarios.
For comparison purposes, we also deploy MADDPG \cite{ddpg} and MATD3 \cite{td3} algorithms within the proposed MADRL framework. Additionally, we compare the scenario where STARS is used only in the refraction mode (referred to as`` w/o reflection''), where the signals are solely reflected by the vehicle body back to the RSU and propagate through STARS to the in-vehicle user. To facilitate a direct performance comparison, we employ $\widetilde{\gamma}^{\mathrm{RSU}} = \frac{1}{N}\sum_{n=0}^{N}\gamma_n^{\mathrm{RSU}}$ to evaluate the overall sensing performance and $\widetilde{R} = \frac{1}{N}\sum_{n=0}^{N}R_n$ to evaluate the overall communication performance. Figs. \ref{f3a} and \ref{f3b} respectively illustrate the S\&C performance versus the number of STARS elements. Firstly, when STARS is in the w/o reflection mode, the performance differences among the deployed algorithms are not very pronounced. However, enhancing reflection through STARS significantly improves sensing performance of $150\%-350\%$, thereby further achieving a 150\% improvement in achievable transmission rate.
It demonstrates the feasibility and importance of jointly improving S\&C to realize superior sensing-assisted communication in target-mounted STARS systems. Secondly, as the number of STARS elements increases, both S\&C performance improves significantly when STARS is in the ``w/ reflection'' mode. Furthermore, two proposed MADRL algorithms demonstrate more significant performance enhancements than the deterministic policies (MADDPG and MATD3), where the poor exploration capabilities make it challenging to reach optimal solutions. Finally, it can be seen that our proposed MADRL framework enables agents to achieve the trade-off between S\&C and the equilibrium cooperation and competition between agents for all algorithms.

Figs. \ref{f4} and \ref{f5} display the S\&C performance for different RSU transmit antennas and receive antennas. Firstly, the simulation results illustrate that both agents can maintain appropriate decisions under various output dimensions of DNNs, demonstrating the stability of the proposed MDP structure within the MADRL framework. From Fig. \ref{f4}, we observe that increasing the number of transmit antennas $N_{\mathrm{t}}$ can effectively enhance S\&C performance, providing additional gains for agents under the relationships of cooperation and competition. In Fig. \ref{f5}(a), it is noticed that an increased number of receive antennas enhances sensing performance owing to better receive filtering ability. The improved communication performance shown in Fig. \ref{f5}(b) also demonstrates the effective sensing-assisted communication scheme in the target-mounted STARS system. Moreover, the two proposed algorithms still perform better than competitors of deterministic strategies.
\begin{figure}[t]
\centering
\includegraphics[width = 3.3 in]{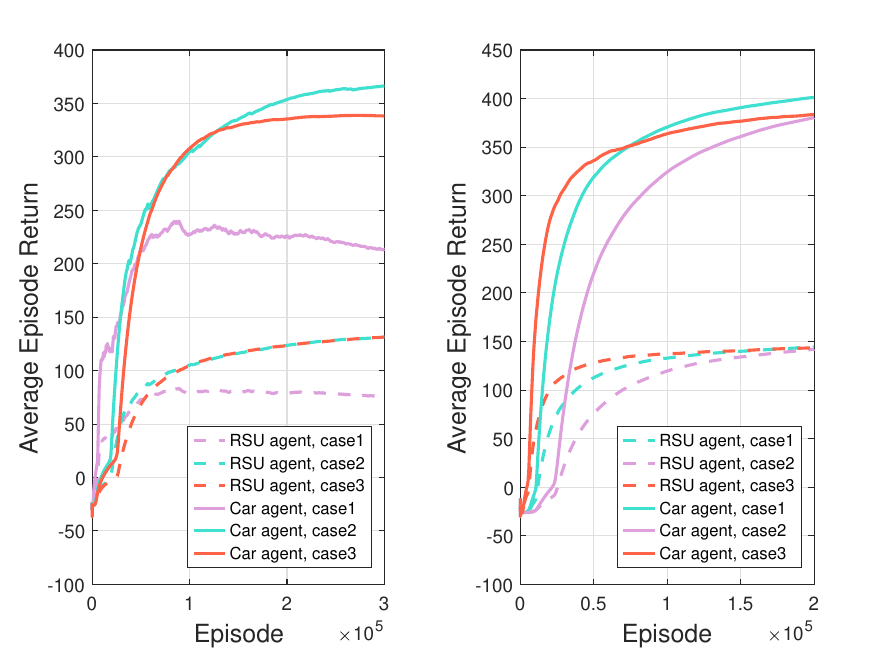}
\footnotesize \hbox{(a) Off-policy MASAC Algorithm. \hspace{0.1cm} (b) On-policy MAPPO Algorithm.}
\caption{The performance of two designed algorithms using different MADRL hyperparameter cases.}
\label{f8}
\vspace{-0.3 cm}
\end{figure}

\begin{figure}[t]
\centering
\includegraphics[width = 3.3 in]{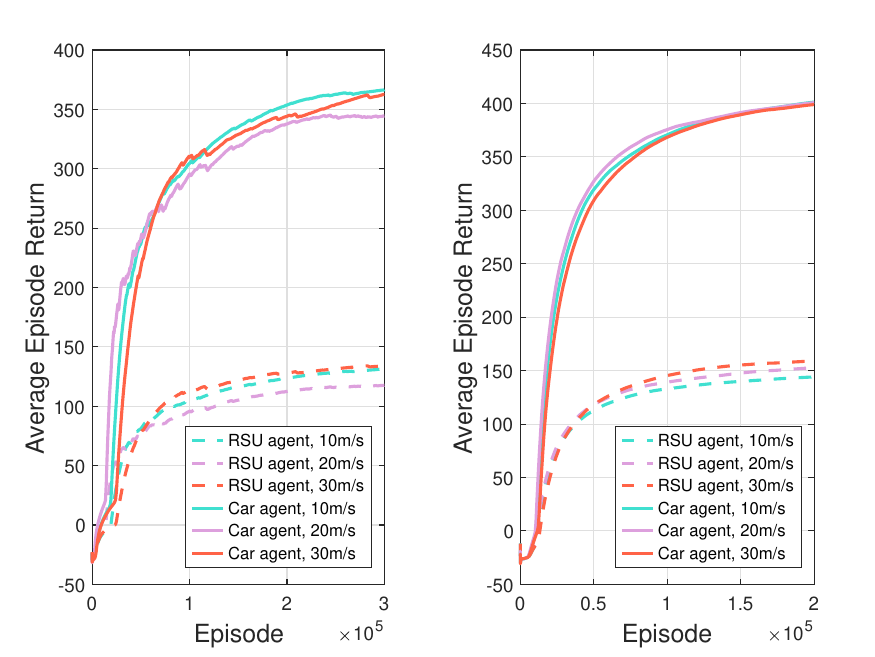}
\footnotesize \hbox{(a) Off-policy MASAC Algorithm. \hspace{0.1cm} (b) On-policy MAPPO Algorithm.}
\caption{The performance of two designed algorithms under different vehicle velocities.}
\label{f9}
\vspace{-0.3 cm}
\end{figure}

We notice from Figs. \ref{f3a}-\ref{f5} that the MASAC algorithm outperforms the MAPPO algorithm in most cases. This is because the off-policy strategy of MASAC ensures the full utilization of all historical experiences. At the same time, the on-policy MAPPO learns from data consistently using the current policy, resulting in lower data utilization and insufficient exploration. However, we should emphasize that the thorough exploration of off-policy algorithms often comes at the cost of increased interaction and learning overhead. In particular, MASAC requires more extensive learning interactions compared with MAPPO. To verify this fact, Fig. \ref{f6} illustrates the average number of episodes required to converge four algorithms under various scenarios. These simulation results indicate that the on-policy MAPPO algorithm can rapidly adapt to unknown environments and requires a fewer number of episodes to achieve convergence.
In contrast, although the incorporation of entropy in MASAC has been shown to enhance exploration capability and learning efficiency, it remains necessary for MASAC to learn extensive knowledge from historical experiences.

Fig. \ref{f7} illustrates the S\&C performance under different transmit power $P$ with $M=64$, $N_{\mathrm{t}}=N_{\mathrm{r}}=36$.
Simulation results demonstrate that our proposed algorithm can attain consistent  S\&C gain despite varying degrees of trade-offs induced by different powers. Moreover, even in challenging scenarios with low power levels, our two proposed algorithms can effectively improve S\&C performance by employing appropriate reward functions \eqref{p15} and \eqref{p16}. It is worth noting that MAPPO adopts the default hyperparameters as shown in TABLE \ref{table:1} for different power levels, while the performance optimization of the other three algorithms requires significant adjustments to the hyperparameters.

Then, in Fig. \ref{f8} we further verify the robustness of proposed algorithms by showing the performance of MAPPO and MASAC under three sets of different hyperparameter configurations (case 1: $\beta_{\phi^{\mathrm{i}}} = \beta_{\theta^{\mathrm{i}}}=2\times 10^{-4}$, case 2: $\beta_{\phi^{\mathrm{i}}} = \beta_{\theta^{\mathrm{i}}}=5\times 10^{-5}$, case 3: $T_{\mathrm{batch}}$ of MASAC $=32$, $T_{\mathrm{max}}$ of MAPPO $=32$). The simulation results reveal that the on-policy MAPPO algorithm maintains stable performance across a certain range of hyperparameter variations, consistently exhibiting superior performance. In contrast, the off-policy MASAC algorithm is highly sensitive to the specified hyperparameters. The instability of MASAC undoubtedly increases the deployment difficulty in real-world environments, as careful tuning of hyperparameters is required for different scenarios. With its strong stability and robustness in various complex environments, MAPPO demonstrates a greater suitability for the target-mounted STARS vehicle network.

Fig. \ref{f9}  illustrates the performance of two algorithms under different vehicle speeds $v_y=10$m/s, 20m/s, and 30m/s. The simulation results demonstrate that both algorithms can effectively facilitate agent learning in environments with different vehicle speeds, even in scenarios with unknown road conditions and CSI. This result highlights the effectiveness of MADRL algorithms in tackling complex problems when agents have access to limited local information.

\begin{figure}[t]
\centering
\includegraphics[width = 3.3 in]{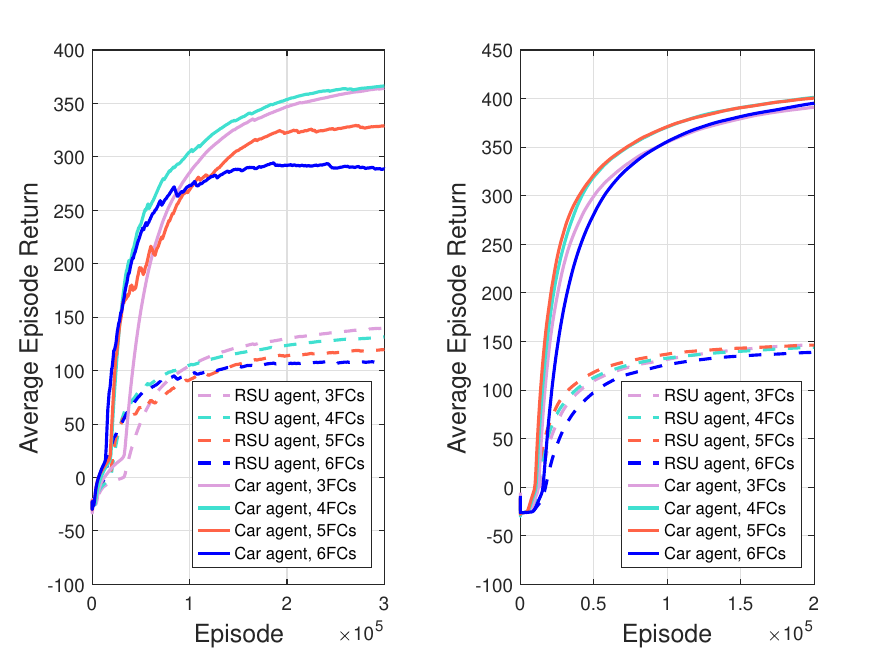}
\footnotesize \hbox{(a) Off-policy MASAC Algorithm. \hspace{0.1cm} (b) On-policy MAPPO Algorithm.}
\caption{The performance of two designed algorithms under different DNN structures.}
\label{f10}
\vspace{-0.3 cm}
\end{figure}

Finally, we compare the impact of the DNN architecture employed by agents. To ensure a fair comparison, each agent consists of fully connected layers (FCs) with 200-width of each hidden layer. Fig. \ref{f10} illustrates the convergence performance of the two algorithms with different numbers of FCs deployed in the actor/critic network, which demonstrates that we can employ lightweight networks (3-4 FCs) to realize S\&C optimization, further verifying the significant advantage of MADRL in tackling continuous decision-making tasks. Moreover, the issue of overfitting caused by gradient invariance becomes severe in MASAC as FCs deepen. In contrast, MAPPO maintains stable performance for networks with different layers and exhibits better overall robustness.


\section{Conclusions}\label{section:G}
In this paper, we proposed an end-to-end MADRL framework for simultaneously enhancing S\&C performance in the target-mounted STARS-assisted ISAC system. A joint design problem was formulated in order to optimize the transmit beamforming and receive filter at the RSU agent, as well as the configuration matrices of STARS at the Car agent.
The simulation results demonstrated that utilizing target-mounted STARS can notably improve the achievable transmission rate for the in-vehicle user.
This improvement is achieved not only by utilizing STARS' refraction property to create a favorable transmission link, but also by using its reflection ability to enhance the echo at RSU, which in turn enhances the radar SNR and enables more accurate beamforming prediction for transmission.
Moreover, we compared and analyzed the robustness and performance of the proposed MASAC and MAPPO algorithms, revealing their advantages compared with deterministic strategies.
The extensive simulation results demonstrate the important role of STARS on the target vehicle and the proposed MADRL algorithm in enhancing sensing-assisted communications for vehicle networks.

\end{document}